\documentclass[%
10pt,twocolumn,superscriptaddress,
 amsmath,amssymb,
 aps,
]{revtex4-2}
\usepackage{epsfig,fullpage,verbatim,amsmath,amsfonts,enumerate,amssymb,graphicx,braket,physics,empheq,subfigure,subcaption,xcolor}
\usepackage{graphicx}
\usepackage{dcolumn}
\usepackage{bm}
\usepackage[english]{babel}
\usepackage[font=footnotesize,labelfont=bf]{caption}
\usepackage{CJKutf8}

\usepackage[colorlinks]{hyperref} 

\begin{document}

\title{An Approach to Probing Particles and Quasi-particles in the Condensed Bose-Hubbard Model}

\author{Huy Nguyen}
\affiliation{
 Joint Center for Quantum Information and Computer Science, University of Maryland-NIST, College Park, Maryland, MD 20742, USA
}
\affiliation{
 Joint Quantum Institute, University of Maryland-NIST,University of Maryland-NIST, College Park, Maryland, MD 20742, USA
}

\author{\begin{CJK}{UTF8}{gbsn}Yu-Xin Wang (王语馨)\end{CJK}}
\affiliation{
 Joint Center for Quantum Information and Computer Science, University of Maryland-NIST, College Park, Maryland, MD 20742, USA
}
\author{Jacob M. Taylor}
\affiliation{
 Joint Center for Quantum Information and Computer Science, University of Maryland-NIST, College Park, Maryland, MD 20742, USA
}
\affiliation{
 Joint Quantum Institute, University of Maryland-NIST,University of Maryland-NIST, College Park, Maryland, MD 20742, USA
}

\date{\today}

\begin{abstract}
\textbf{Abstract:} Measurement plays a crucial role in a quantum system beyond just learning about the system state: it changes the post-measurement state and hence influences the subsequent time evolution; further, measurement can even create entanglement in the post-measurement conditional state.
In this work, we study how careful choice of parameters for a typical measurement process on cold atoms systems -- phase contrast imaging -- has a strong impact on both what the experimentalist observes but also on the backaction the measurement has on the system, including the creation and diffusion of quasiparticles emerging from the quantum many-body dynamics. We focus on the case of a Bose-Einstein-condensate array, in the low-temperature and low-momentum limit. Our theoretical investigation reveals regimes where the imaging light probes either the bare particle or quasiparticle dynamics. Moreover, we find a path to selectively measuring quasiparticle modes directly, as well as controlling over the measurement-induced creation and diffusion of quasiparticles into different momentum states. This lays a foundation for understanding the effects of both experimental approaches for probing many-body systems, but also more speculative directions such as observable consequences of `spontaneous collapse' predictions from novel models of quantum gravity on aspects of the Standard Model.
\end{abstract}

\maketitle

\section{Introduction}
Quantum many-body systems display rich, emergent dynamics when at low temperatures or near their quantum ground state, and have been the subject of nearly 100 years of experimental and theoretical study. One particular domain of successful theory-experiment connection has been in cold atom experiments~\cite{bloch_many-body_2008}, where the ultra-cold atomic gases and their interactions have led to the observation of a Bose-Einstein condensate (BEC)~\cite{anderson_observation_1995}, vortex lattices~\cite{abo-shaeer_observation_2001}, superfluid-Mott Insulator transitions~\cite{greiner_quantum_2002}, fermionic superfluidity~\cite{regal_observation_2004,zwierlein_condensation_2004}, and more exotic phases~\cite{fallani_ultracold_2007}. More recently, theoretical and experimental efforts have started to examine how continuous or stroboscopic measurement of systems can lead to phase transitions and other dramatic changes in their dynamics~\cite{hurst_feedback_2020}. Specifically, the high control and low entropy of these experimental systems mean they are often deeply in the quantum domain, and thus are amenable to being a testbed for components of quantum theory, including measurement.

Here we explore the specific question of how a measurement apparatus often used in the field -- phase contrast imaging \cite{zernike_phase_1942} -- translates to observations on the many-body system. We are particularly interested in the interplay between emergent dynamics, as described by quasi-particles~\cite{bogolubov_theory_1971}, and the underlying physical system comprising individual atoms.   Already many groups have observed that measurement of these many-body systems leads to heating due to the creation of quasi-particles~\cite{altuntas_weak-measurement-induced_2023}. 
By careful examination of the amplitude and character of the measurement, we find that the same phase-contrast-imaging setup can be used in one mode for detecting quasi-particles without adding much heating, and in another mode to directly image the atoms, which naturally adds quasiparticles to the condensate.

This approach has wide-ranging implications for both experimental quantum measurement techniques as well as questions of fundamental relevance to physics. 
On the practical side, it may enable more efficient use of laser imaging systems for investigating specific quantum many-body systems. Furthermore, on the theoretical side, this approach provides a foundation for exploring the role of renormalization in quantum field theories and its interplay with how such theories are probed. A quantitative understanding of this interplay goes beyond the usual qualitative commentary that the energy you probe at is the energy of your renormalization cutoff.  Of particular relevance are scenarios, such as in gravitationally-induced decoherence \cite{anastopoulos_master_2013} or theories of spontaneous collapse \cite{diosi_universal_1987}, where there may be observable consequences of the effective measurement these decoherence mechanisms create that our theory helps to provide a tool to calculate. Thus, our work lays the path for future efforts to fully explore this novel domain of understanding. 

The paper is structured as follows: Sec.~\ref{sec:level1} begins with a toy model of an atom in a double-well potential under local weak continuous measurement, outlining the general formalism and demonstrating how effective observations emerge by tuning measurement bandwidths in two opposite regimes. Section~\ref{sec:level4} then extends this setup to a BEC array, modeled by a weakly interacting Bose-Hubbard Hamiltonian, in which the measurement procedure in the wide (narrow) bandwidth yields effective observations of bare particles (Bogoliubov quasiparticles). In Sec.~\ref{sec:level5}, we show that the observation of bare particles results in significant quasiparticle heating in the condensate, while observing the quasiparticles does not. We conclude in Sec.~\ref{sec:level6} with an outlook.

\section{\label{sec:level1} A weakly Measured Atom in A Double-well Potential}

\begin{figure}[h]
\centering
    
    \begin{subfigure}
        \centering
        \includegraphics[width=0.95\columnwidth ]{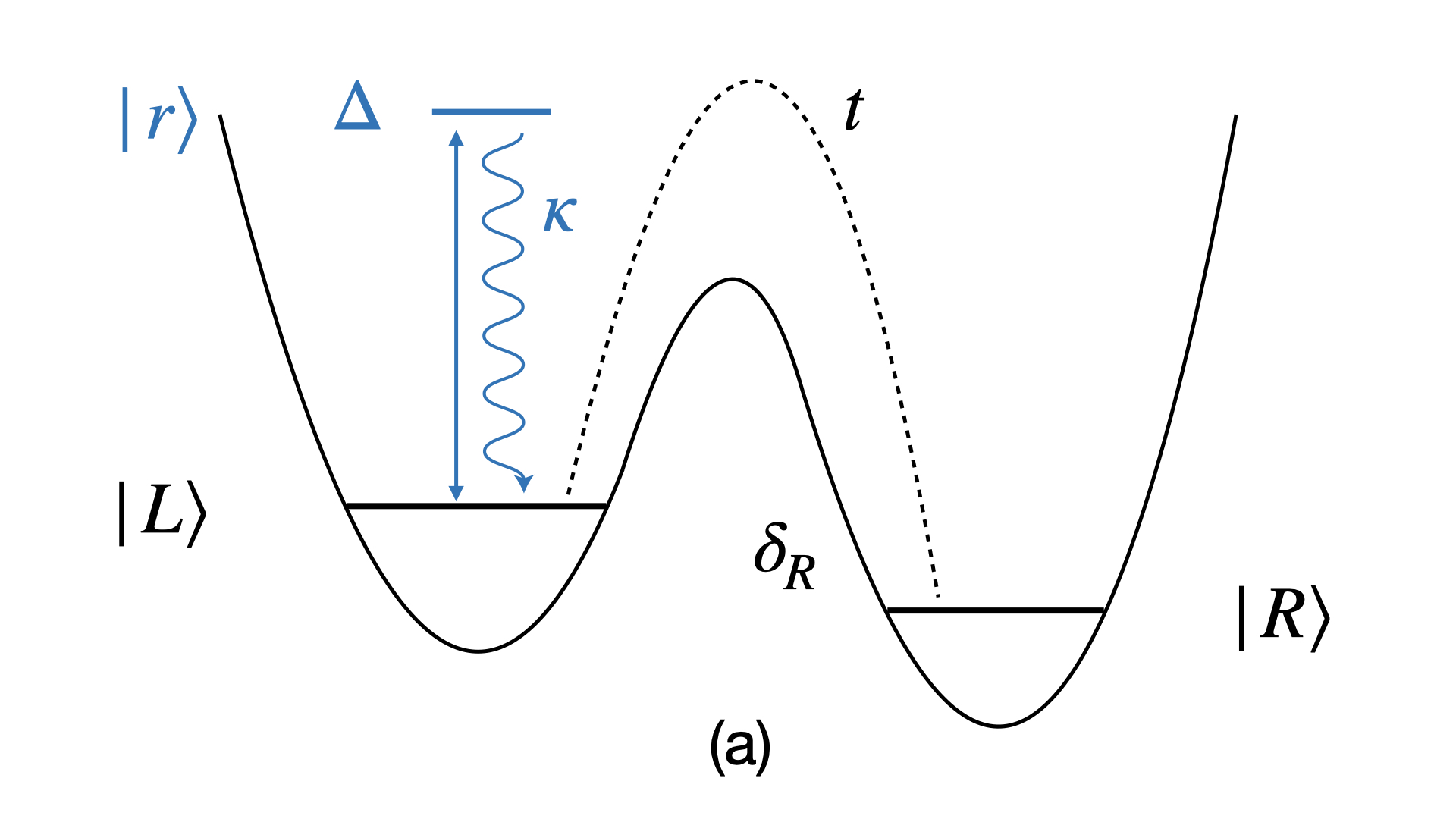}
        \label{fig:sub1}
    \end{subfigure}
    
    
    \begin{subfigure}
        \centering
        \includegraphics[width=7cm]{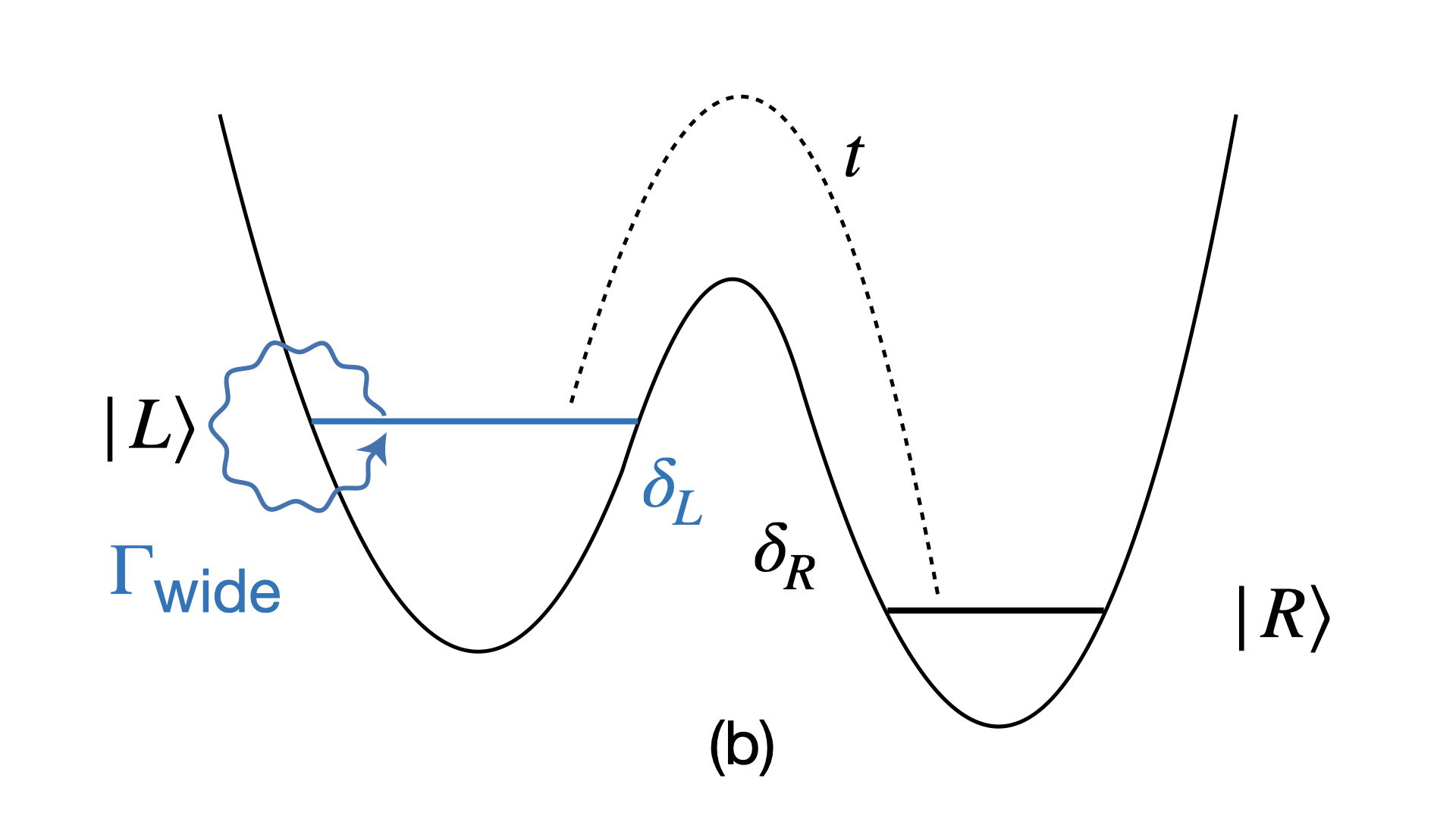}
        \label{fig:sub2}
    \end{subfigure}
    
    
    \begin{subfigure}
        \centering
        \includegraphics[width=7cm]{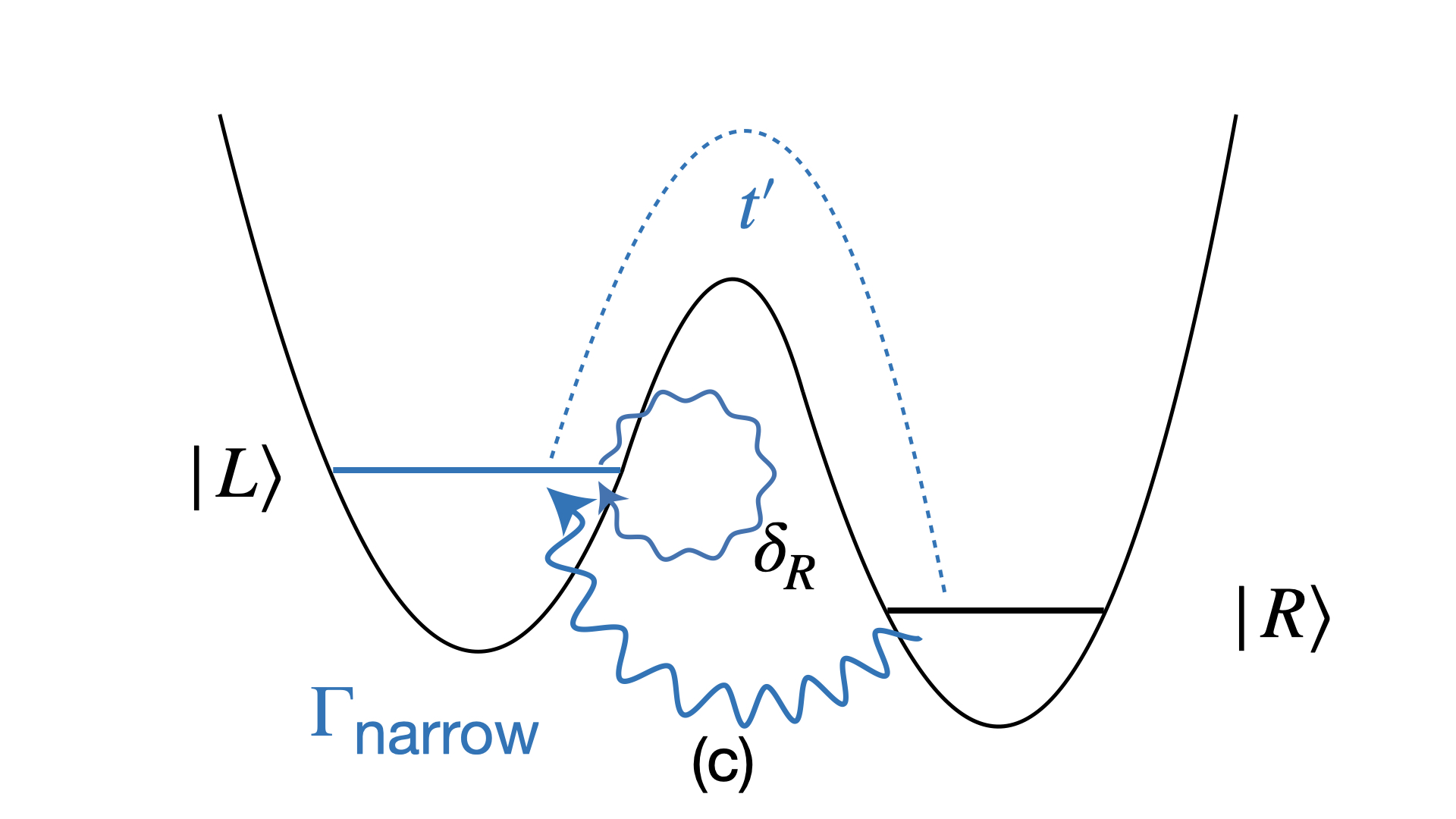}
        \label{fig:sub3}
    \end{subfigure}
\caption{\label{fig:2level} 
(a) Schematic of a weakly probed atom in a double-well potential. 
(b) The effective tunneling atom in the wide-bandwidth regime. Adiabatic elimination of the fast probe state only introduces a Stark-like detuning in $\ket{L}$ (or equivalently a Stark-like shift to $\delta_R$ of $\ket{R}$, as explained below), while retaining the tunneling rate $t$ between the double-well states. Notice that the effective shelving jump operator measures the probability of the atom being in the state $\ket{L}$. 
(c) The effective tunneling atom in the narrow bandwidth. Adiabatic elimination of the now dynamical probe state introduces both a detuning at $\ket{L}$ and a shift to the tunneling rate $t$, as a result of perturbatively absorbing and eliminating the probe state interaction. Notice that the jump operator is a superposition of shelving at $\ket{L}$ and driving a transition from $\ket{R}$ top $\ket{L}$, which means measuring a superposition of $\ket{R}$ and $\ket{L}$.}
\end{figure}

In this section, we consider an atom in a double-well potential under weak continuous measurement. In the left well, a coherent drive allows the atom to Rabi-flop between $\ket{L}$ and a metastable internal state. This internal state is in turn subject to weak and continuous measurements. Adiabatically eliminating the internal state thus enables the extraction of the effective measurement-induced dynamics of the tunneling atom. Through this example, we establish the general procedure of extracting the effective measurement observables and the measurement-induced changes to the system Hamiltonian (also known as the Lamb shift). 

We model an atom in a double-well potential, with trapped states $\ket{L}$ and $\ket{R}$, and a probe state $\ket{r}$ by the Hamiltonian $H = H_{\text{atom}} + V$, where 
\begin{align}
H_{\text{atom}}&=\delta_R\sigma_{RR}-t\left(\sigma_{LR}+\sigma_{RL}\right),\\
    V&=\Omega\left(\sigma_{Lr}+\sigma_{rL}\right)
.
\end{align}
Here, we define $\sigma_{ij}\equiv\ket{i}\bra{j}$ for $i,j=L,R,r$, and the right-hand-side trap potential state $\ket{R}$ is detuned at $\delta_R$ and the probe state $\ket{r}$ at $\Delta$. The tunneling rate within the potential, and the Rabi frequency between the atom state $\ket{L}$ and probe state $\ket{r}$ are $t$ and $\Omega$, respectively. A laser beam continuously monitors the probe state $\ket{r}$, which is metastable. When averaging over all possible measurement outcomes, this measurement induces a relaxation process from $\ket{r}$ to $\ket{L}$.
We denote by $\Gamma $ the jump operator describing this dissipative process:
\begin{align}
\label{eq:2level.Gamma}
    \Gamma &=\sqrt{\kappa}\sigma_{Lr}
    .
\end{align}
Figure~\ref{fig:2level}(a) illustrates the atomic levels and the dissipative process.

The expectation values of $\sigma_{ij}$ evolve under the adjoint Lindblad super-operator $\mathcal{L}^\dagger$, 
which reads
\begin{align}
    &\frac{d}{dt}\langle \sigma _{ij} (t) \rangle{}
    =\text{Tr}(\sigma_{ij}\dot{\rho} (t))=\text{Tr}\left[(\mathcal{L}^\dagger\sigma_{ij})\rho\right],\\
    \mathcal{L}^\dagger\sigma_{ij}&=i[H,\sigma_{ij}]+\underbrace{\frac{1}{2}\left(2\Gamma^\dagger\sigma_{ij}\Gamma-\{\Gamma^\dagger\Gamma,\sigma_{ij}\}\right)}_{\mathcal{L}^\dagger_{\text{diss}}[\Gamma]\sigma_{ij}}.
    \label{eq:qme.adjoint}
\end{align}

\begin{figure}[h]
\includegraphics[width=8cm]{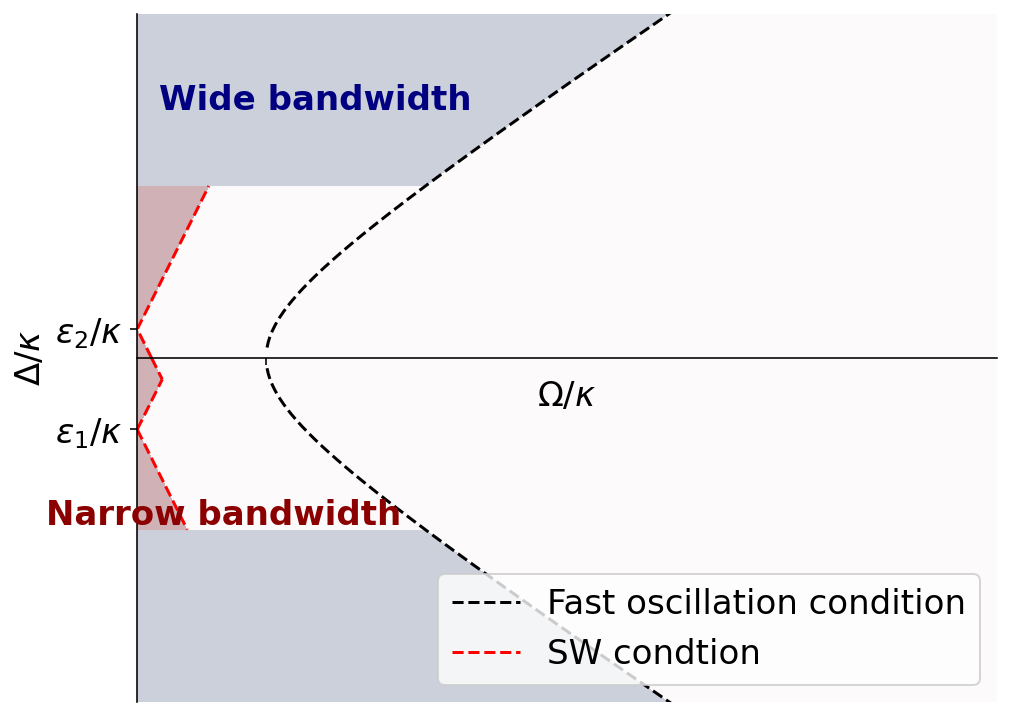}
\caption{\label{fig:bandwidth} A diagram describing the measurement bandwidths of interest ($\kappa$ fixed). In the wide bandwidth regime, the tunneling atom does not ``see" the drive between $\ket{L}$ and $\ket{r}$ due to the large detuning $\Delta$ and damping rate $\kappa$, but the drive strength $\Omega$ is still comparable to the energetic parameters of the atom. In the narrow bandwidth regime, the detuning and damping are not necessarily large, but the drive strength $\Omega$ is small enough to act perturbatively on the atom.}
\end{figure}

We are interested in two regimes: (1) where the internal $\ket{r}$ state dynamics is much faster than that of the two-level atom, so that we can trace out the dynamics of $\ket{r}$ and obtain an effective measurement observable as well as an effective system Hamiltonian, and (2) where the internal $\ket{r}$ state is relatively dynamical but the associated laser drive $\Omega$ is small enough to be treated perturbatively. In what follows, we perform this analysis in two different regimes of the measurement bandwidths. 
We illustrate these regimes of measurement bandwidths in Fig.~\ref{fig:bandwidth}. In the \textit{wide} measurement-bandwidth regime, the detuning $\Delta$ of the probe state is much larger than the Rabi drive $\Omega$, while 
the strengths of intrinsic parameters of the tunneling system, $\delta_R$ and $t$ are comparable to $\Omega$. Additionally, the damping rate $\kappa$ is taken to be large relative to $\Omega$, implying that the dynamical variables of $\ket{r}$, namely, $\sigma_{ri}$ ($i=0,1,e$) evolve and decay much more rapidly than those that only involve the trap potential levels $\ket{L},\ket{R}$. 
In this regime, the dynamics of $\ket{r}$ are approximately determined by observables that act only on the tunneling system, i.e., by $\sigma_{LL},\sigma_{LR},\sigma_{RR}$. As we show in Sec.~\ref{sec:level2}, we can use the standard adiabatic elimination approximation (see, e.g., \cite{reiter_effective_2012}) to integrate out fast-evolving degrees of freedom and obtain a measurement-induced Stark-like shift in the effective atomic frequency, as well as an effective measurement observable formed by the bare atomic states.

On the other hand, in the \textit{narrow} measurement-bandwidth regime, the Rabi drive $\Omega$ and damping rate $\kappa$ are small in comparison to system parameters $t,\delta_R$, and $\Delta$. Physically, this means that coupling interaction $V$ between $\ket{r}$ and $\ket{L}$ only introduces a weak perturbation to the two-level Hamiltonian. 
We account for system dynamics due to this weak perturbation through the Schrieffer-Wolff (SW) transformation.
Under the SW transformation, the ``dressed" atom and the probe state become decoupled, accompanied by second-order in $\Omega$ corrections to the external state detuning and the internal Hamiltonian of the atom in the double-well potential, as well as an $O(\Omega)$ correction $\delta\Gamma$
to the jump operator. We shall see in Sec.~\ref{sec:level3} that this $\delta\Gamma$ is the leading order contribution in the effective jump acting directly on the atom in the double well, which effectively describes a measurement process with respect to a superposition state between the atomic trapped states in the well.

\subsection{\label{sec:level2}Wide measurement-bandwidth regime}
Upon adiabatic elimination of the fast-evolving external state $\ket{e}$, the effective adjoint Lindbladian  describing the time evolution of the two-level atom is uniquely specified by the following equations of motion
\begin{align}
    \mathcal{L^\dagger}\sigma_{LL} &=i[H_{\text{atom}},\sigma_{LL}],\\
    \mathcal{L^\dagger}\sigma_{RL} &=i[H_{\text{{atom}}},\sigma_{RL}]+i\delta_L\sigma_{RL}-\frac{\kappa_{\text{wide}}}{2}\sigma_{RL},\\
    \mathcal{L}^\dagger\sigma_{RR} &=i[H_{\text{atom}},\sigma_{RR}],
\end{align}
where we define $\delta_L \equiv\frac{\Omega^2}{(\Delta-\delta_R)^2+\kappa^2/4}(\Delta-\delta_R)$ and $\kappa_{\text{wide}}\equiv\frac{\Omega^2}{(\Delta-\delta_R)^2+\kappa^2/4}\kappa$. Comparing above equations of motion to that of the bare tunneling atom given by $H_{\text{atom}}$, we see that the measurement induces two changes to the atom dynamics: it introduces a new dissipative term acting on the transition variables $\sigma_{RL}$ and $\sigma_{LR}$, as well as a Hamiltonian correction whose strength is set by $\delta'_0-\delta_R$, which can be viewed as a Stark-like shift to the atom frequency.
We thus obtain the effective system Lindbladian of the form in Eq.~\eqref{eq:qme.adjoint}, with the effective two-level Hamiltonian and jump operator given by
\begin{align}
    H_{\text{eff,wide}} &= H_{\text{atom}}+\delta_L\sigma_{LL},\\
    \Gamma_{\text{eff,wide}}&=\sqrt{\kappa_{\text{wide}}}\sigma_{LL}.
\end{align}
This shows that, when driving the transition from the probe state $\ket{r}$ to $\ket{L}$ in the wide-bandwidth regime, we are effectively measuring the probability of the atom in the bare atomic states $\ket{L}$ (or equivalently $\ket{R}$) at rate $\kappa_{\text{wide}}$, while introducing a shift in the atom frequency $ \delta_L$. 

We emphasize that when applying the adiabatic elimination approximation and setting time-derivatives of $\sigma_{rj}$ (for $j=R,L$) to zero, we assume that the tunneling rate $t$ is negligible compared with dynamics of the external level. This ensures that the \textit{fast} dynamics of $\sigma_{rj}$ does not ``see" the \textit{slow} transition terms associated with the evolutions of the state $\ket{L},\ket{R}$. 
In contrast, if $t$ is not small,
the time-evolution of $\sigma_{Lr},\sigma_{Rr}$ would be comparable to that of $\ket{L},\ket{R}$, thereby invalidating the separation of timescales required for the adiabatic elimination to hold.

\subsection{\label{sec:level3}Narrow measurement-bandwidth regime}
In the narrow bandwidth regime, the weak coupling interaction $V=\Omega(\sigma_{Lr}+\sigma_{rL})$ serves as a perturbation to the net system, and we move to a dressed basis using a SW transformation $e^S$ that decouples the two atomic states $\ket{L},\ket{R}$ from the probe state $\ket{r}$. The generator $S$ for our system is given by,
    \begin{equation}\label{2levelSW}\small
        S = \frac{\Omega}{\Omega_{SW}^2}\left[-t(\sigma_{rR}-\sigma_{Rr})+(\Delta - \delta_R)(\sigma_{rL}-\sigma_{Lr})\right],
    \end{equation}
which satisfies the condition $V+[S,H_{\text{atom}}+\Delta\sigma_{rr}]=0$, and we define $\Omega_{\text{SW}}\equiv\sqrt{\Delta^2-\delta_R\Delta-t^2}$. This is valid as long as the Rabi frequency $\Omega$ of the weak drive $V$ is a lot smaller than the energy gap between subspaces of the two-level atom and the external state $\ket{r}$, i.e $\Omega\ll|\Delta-\epsilon_1|,|\Delta-\epsilon_2|$, where $\epsilon_{1,2}\equiv\frac{1}{2}\left(\delta_R\mp\sqrt{\delta_R^2+4t^2}\right)$ are the eigenenergies of the tunneling atom in the absence of coupling to the external level.
This method (i.e., second-order SW transformation) is akin to the typical second-order perturbation theory. As mentioned, the transformation $e^SHe^{-S}$ serves to decouple the probe state $\ket{r}$ from the system states $\ket{L}$ and $\ket{R}$ in the system Hamiltonian. However, at the Hamiltonian level, this comes at a cost of perturbatively dressing the Rabi oscillations between the two atomic levels and creating a detuning at state $\ket{L}$. In the dissipative action, the jump operator, under the $S$ transformation, now also acts perturbatively on the atomic two level states. We then apply adiabatic elimination of the probe state $\ket{r}$ to the total system master equation under these perturbed Hamiltonian and jump operator, while maintaining the leading order $O(\Omega^2)$ term throughout the calculation. Similar to the \textit{wide}-bandwidth case, 
we obtain an effective Hamiltonian and jump operator acting solely on the atom states $\ket{L}$ and $\ket{R}$ (for calculation details, see App.~\ref{apppendix:a}):
\begin{align}
    H_{\text{eff,narrow}}&=H_{\text{atom}}+\delta_L'\sigma_{LL}\nonumber\\
    &+\frac{\Omega^2 t}{2\Omega_{SW}^2}(\sigma_{LR}+\sigma_{RL}),\\
    \Gamma_{\text{eff,narrow}} &=\sqrt{\frac{t^2\Omega^2}{\Omega_{SW}^4}\kappa}\left(\sigma_{LR}+\frac{\delta_R-\Delta}{t}\sigma_{LL}\right)&\label{Gamma_single_narrow}
    .
\end{align}

In the effective tunneling Hamiltonian under narrow bandwidth, the "dressed" tunneling rate $t'\equiv\left(1-\frac{\Omega^2}{2\Omega_{SW}^2}\right)t$, and the Stark shift $\delta_L'\equiv-\frac{\Omega^2(\Delta-\delta_R)}{\Omega_{SW}^2}$ to the atom frequency originate solely from the Hamiltonian corrections due to the SW transformation. Further, the effective jump operator in Eq.(\ref{Gamma_single_narrow}) implies an effective transition to $\ket{L}$, from the following superposition of the double-well states,
\begin{equation}
    \ket{\psi_{\text{narrow}}}\propto\ket{R}+\frac{\delta_R-\Delta}{t}\ket{L}.
\end{equation}
This means that the measured state is a superposition determined by the detuning $\Delta$, allowing selective measurement of \textit{eigenstates} of the tunneling Hamiltonian $H_{\text{atom}}$. For example, if the desired measure state is $\ket{\epsilon_1} \propto \frac{1}{2}\left(\delta_R-\sqrt{\delta_R^2+4t^2}\right)\ket{L} -  t \ket{R}$ with eigenenergy $\epsilon_1=\frac{1}{2}(\delta_R-\sqrt{\delta_R^2+4t^2})$, then we simply tune (near-resonance) at $\Delta= \delta_R + \frac{t^2}{\epsilon_1}$. The protocol remains valid provided the Schrieffer-Wolff condition $\Omega\ll|\Delta-\epsilon_1|,|\Delta-\epsilon_2|$ is satisfied.
\begin{figure}[ht]
\centering
\begin{minipage}{0.8\columnwidth}
  \centering
  \includegraphics[width=\linewidth]{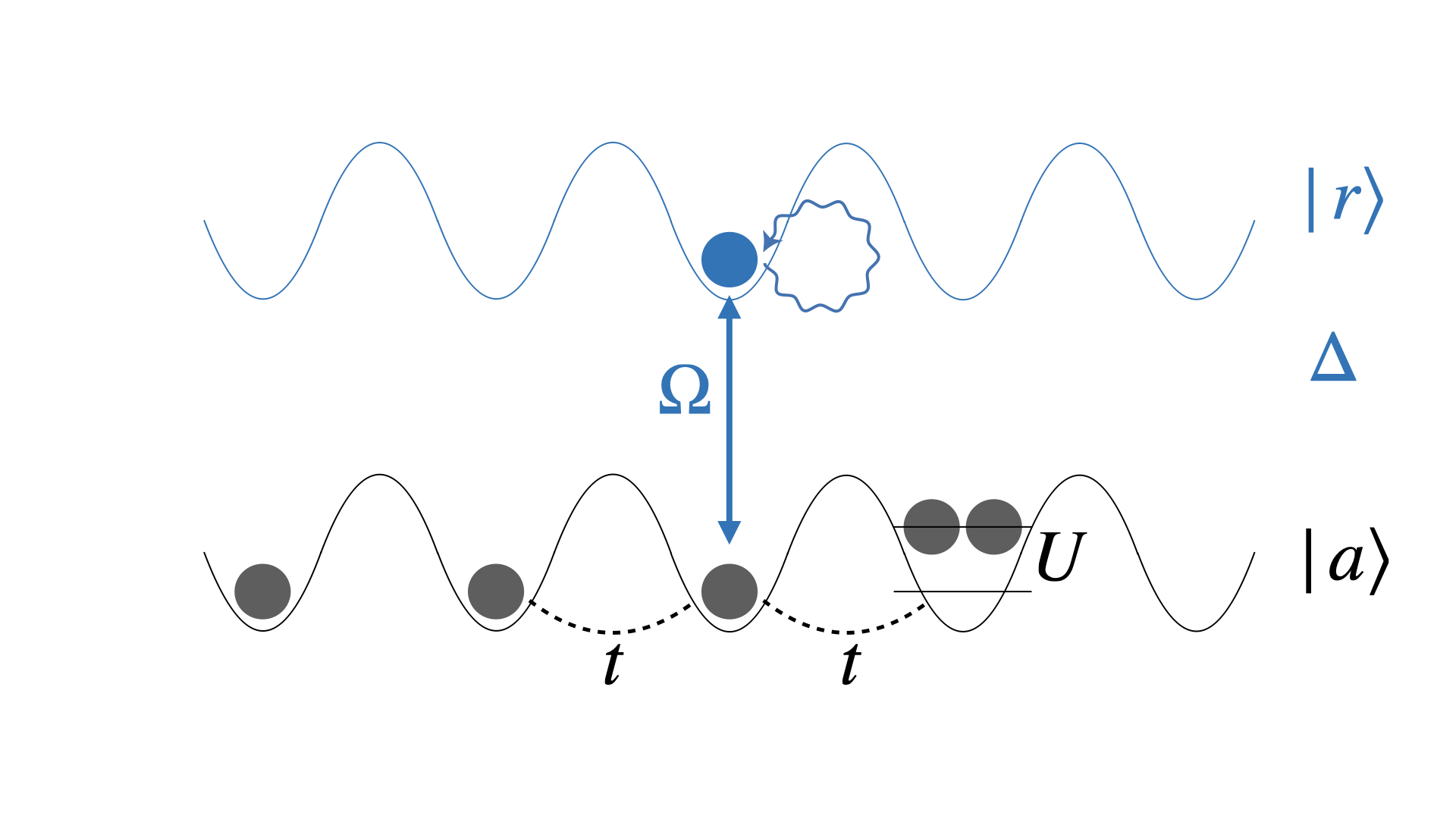}
  \caption*{(a) Particle conserving $\Gamma=\sqrt{\kappa}r^\dagger r$}
\end{minipage}

\vspace{0.5cm}  

\begin{minipage}{0.8\columnwidth}
  \centering
  \includegraphics[width=\linewidth]{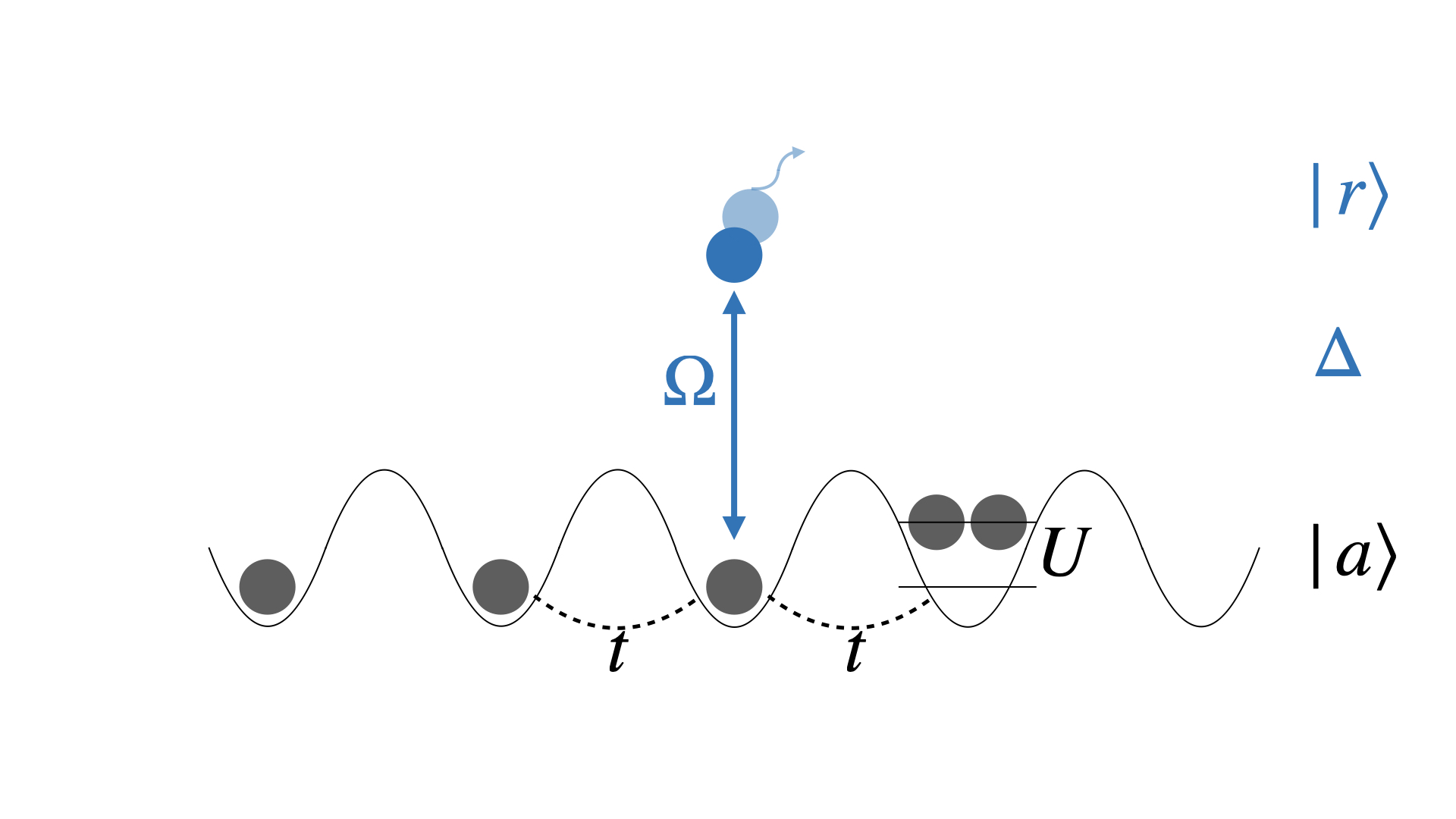}
  \caption*{(b) Particle loss $\Gamma=\sqrt{\kappa}r$}
\end{minipage}

\caption{Weakly-interacting Bose–Hubbard lattice schematic, where the bare bosonic state $a_{j_0}^\dagger\ket{\text{vac}}$ at site $j=j_0$ is Rabi driven to an excited probe state $r^\dagger\ket{\text{vac}}$. (a)~Lossless measurement in which the excited probe bosons remain in the trap. (b) Dissipative measurement in which the excited probe bosons leak from the trap.}
\label{fig:bose_hubbard}
\end{figure}

\section{\label{sec:level4}Bose-Hubbard Condensate under Weak Measurement}
In studying the effect of weak measurement on a single atom tunneling in double-well potential, we have established a baseline understanding of how measurement bandwidth can affect the quantity being probed and a procedure for tuning measurement parameters to target either directly the bare atomic Hamiltonian eigenstates or a desired state superposition. In this section, we now extend the procedure for deriving the effective system Hamiltonian and measurement dynamics to a weakly interacting 1-dimensional Bose-Einstein condensate under weak measurement. Specifically, we consider a 1-dimensional Bose-Hubbard lattice model (engineered via an optical trap set-up), where a Rabi drive is applied to couple a bare bosonic particle state $a_{j_0}^\dag \ket{\text{vac}}$ to an excited state $r^\dag \ket{\text{vac}}$. The excited state is subject to a weak measurement. To connect this model with the single-atom case, we promote the single-particle states to bosonic field operators according to the following mapping:
\begin{align*}
    \ket{r}&\to r^\dagger \ket{\text{vac}}\\
    \ket{L}&\to a_{j=j_0}^\dagger\ket{\text{vac}}\\
    \ket{R}&\to \{a_{j\neq j_0}^\dagger\ket{\text{vac}}\}
    .
\end{align*}

This construction embeds the aforementioned physics of the single-atom-measurement example into an interacting many-body lattice system. We shall also consider two types of measurements: \textit{particle loss} - in which the bosons occupying the excited mode $r^\dag \ket{\text{vac}}$ are untrapped and leak from the system, and \textit{particle counting (lossless)} - in which the probe bosons remain confined.

In general, the introduction of on-site interactions give rise to nonlinear dynamics and a nontrivial interplay with the measurement process. To turn this complex dynamical system into a theoretically tractable problem, and also motivated by practical scenarios involving monitored BECs, in the following discussions, we focus on the case where the system is prepared in a macroscopically large condensate at zero temperature. We further assume that the measurement process is sufficiently slow that the condensate is never destroyed during the time evolution.

This mean-field regime allows us to approximate the weak interaction as a squeezing-type, quadratic interaction—both in real and momentum space—thereby enabling diagonalization of the mean-field Bose-Hubbard Hamiltonian via a linear Bogoliubov transformation.

\subsection{The mean-field Bose-Hubbard Hamiltonian and the Bogoliubov transformation}
We start by providing a brief overview of the Bogoliubov theory. A weakly interacting Bose gas, in a 1D lattice with $N$ sites, can be described by the Bose-Hubbard Hamiltonian
\begin{align}
    H_{BH} &= 
    \sum_{j} \left [ -t\left(a_j^\dagger a_{j+1}+\text{h.c}\right) +\frac{U}{2}n_j(n_j-1)-\mu n_j\right ],
\end{align}
where $a_j$ and $a_j^\dagger$ are the bosonic annihilation and creation operators at the site $j$ satisfying the canonical commutation relation $[a_j,a^\dagger_{j'}]=\delta_{j,j'}$. $t$ is the nearest-neighbor tunneling strength, $\mu$ is the on-site chemical potential and $U$ is the weak on-site repulsive interaction strength. We assume the system state is close to a zero-temperature condensate state in the zero-momentum mode throughout time evolution, so that we can apply mean-field approximation to the Hamiltonian.
That means, suppose $a_{k=0}\approx a^\dagger_{k=0}\approx\sqrt{N_0}$ where the number of bosons at zero momentum $N_0$ is close to $N$ (\textit{Bogoliubov approximation}), we can write $a_j = \sqrt{\frac{N_0}{N}}+\delta a_j$, where $\delta a_j=\frac{1}{\sqrt{N}}\sum_{k\neq0}a_{k}e^{-ijk}$. Prescribing $\delta a_j\to a_j$ such that $a_j$ contains only the non-zero momentum modes, we obtain the mean-field Hamiltonian,
{\small
\begin{align}\label{bose_hubbar_bare}
    H_{\text{MF}} &= -t\sum_{j}(a_j^\dagger a_{j+1}+a^\dagger_{j+1}a_j)+(2Un_0-\mu)\sum_{j}a_j^\dagger a_j\nonumber\\
    &+\frac{Un_0}{2}\sum_j(a_j^2+a^{\dagger2}_j)+E_{0},
\end{align}
}%
where we define the condensate density to be $n_0\equiv\frac{N_0}{N}$, and $E_{0}=(-2t-\mu+Un_0/2)N_0$ is the energy associated with the $k=0$ mode in the condensate. To fix the chemical potential, we ignore the energy contribution and population of the weakly interacting excitations. Therefore, $\mu$ is the energy such that the internal energy of the condensate in $k=0$ mode is minimized with respect to $N_0$, which yields $\mu = -2t+Un_0$.

On the other hand, $H_{MF}$ in the momentum basis takes the form
\begin{equation}
    H_{\text{MF}} = E_{0}+\sum_{k\neq0}\left[\mathcal{A}_k a_k^\dagger a_k+\frac{Un_0}{2}(a_ka_{-k}+a_k^\dagger a_{-k}^\dagger)\right],
\end{equation}
where we define $\mathcal{A}_k\equiv2Un_0-\mu-2t\cos (k)$. The above form of $H_{\text{MF}}$ is diagonal in the Bogoliubov quasiparticle $b_k$ basis, defined by the Bogoliubov transformation
\begin{align}\label{bogo_trans}
    a_k &=u_kb_k-v_kb^\dagger_{-k},\\
    a^\dagger_{k} &=u_kb^\dagger_k-v_kb_{-k}.
\end{align}
Under the Bogoliubov transformation, the diagonal form of $H_{\text{MF}}$ is
\begin{align}
    H_{\text{MF}}  &= \sum_{k\neq0}E_{b,k}b^\dagger_kb_k\nonumber\\
    &\underbrace{+\frac{1}{2}\sum_{k\neq0}(E_k-2Un_0+\mu+2t\cos ak)+E_{0}}_{\text{vacuum dressed by "quantum depletion" }}\nonumber\\
    &= \sum_{k\neq0}E_{b,k}b^\dagger_kb_k+E_{\text{cond}},&\label{H_MF}
\end{align}
where $E_{b,k}\equiv\sqrt{\mathcal{A}_k^2-(Un_0)^2}$ is the Bogoliubov eigenenergy, and $E_{\text{cond}}\equiv\frac{1}{2}\sum_{k\neq0}(E_k-2Un_0+\mu+2t\cos k)+E_{0}$ is the true ground-state energy of the condensate dressed by quantum depletion. The Bogoliubov coefficients are given by
\begin{align}
    u_k&=\sqrt{\frac{1}{2}\left(\frac{\mathcal{A}_k}{E_{b,k}}+1\right)}, \quad 
    v_k=\sqrt{\frac{1}{2}\left(\frac{\mathcal{A}_k}{E_{b,k}}-1\right)}.
\end{align}
In the weakly interacting limit ($U\ll t$), the relative amplitudes of $v_k$ and $u_k$ in the Bogoliubov transformation behave differently for low momenta and high momenta. At low momentum $k$, these coefficients are equal in amplitude, which means that a Bogoliubov quasiparticle is made of the original bare particle propagating along the momentum $k$ and another backward-propagating hole along $-k$. In contrast, at high momentum, $v_k$ vanishes which implies that a Bogoliubov quasiparticle behaves almost like the bare particle. In addition, the value $v_k^2$ is often referred to in some literature as the "quantum depletion" of the condensate, as it is the ground-state value  of $\langle{a_k^\dagger a_k} \rangle$ in each mode $k$.

\begin{figure}[t]
\includegraphics[width=8.8cm]{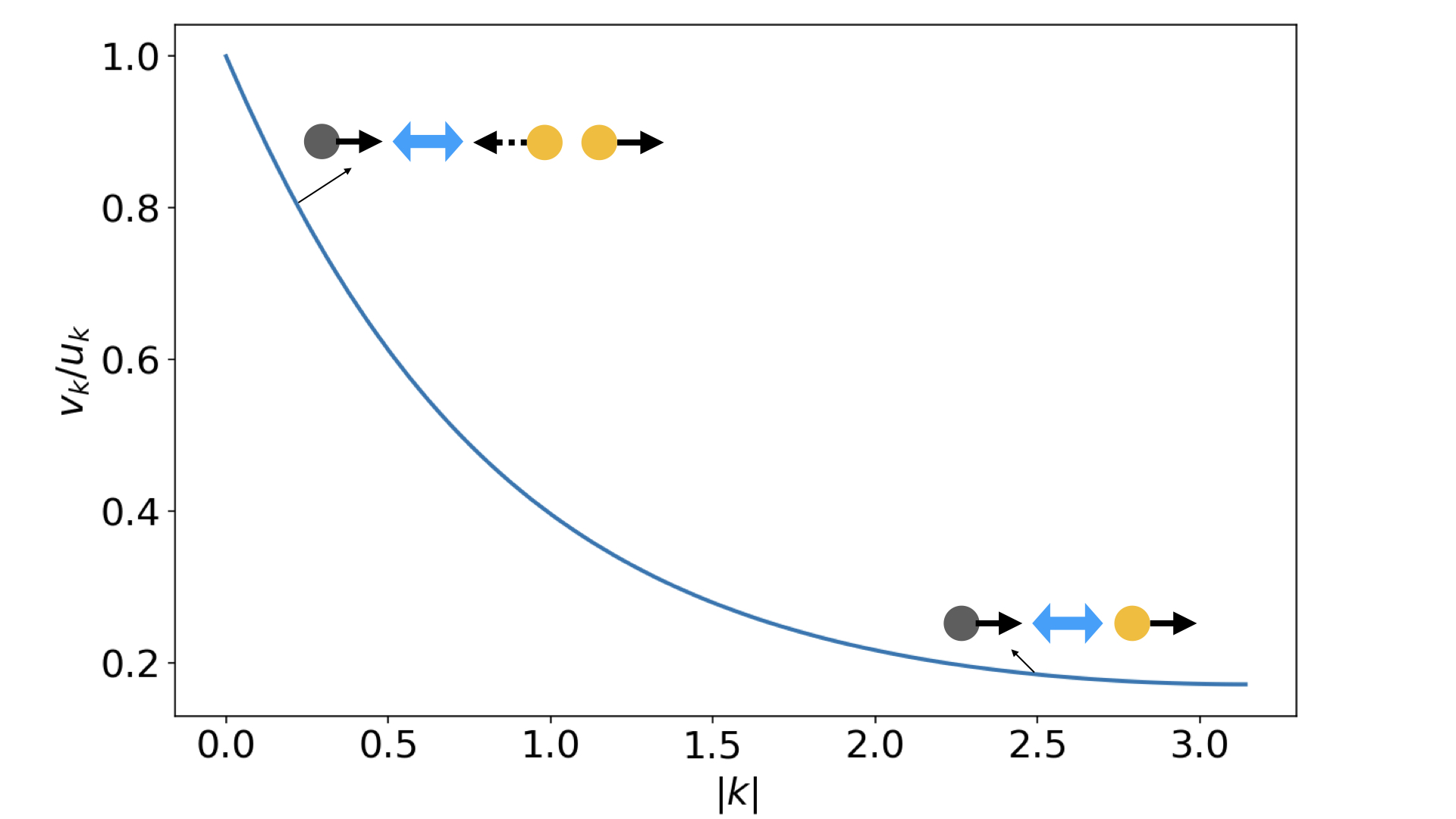}
\caption{\label{fig:bogoliubov} The correspondence between bare bosons (in momentum space) and Bogoliubov quasiparticles, described by the relative ratio between Bogoliubov parameters as function of momentum norm. In higher momenta (or higher excitations), the bosons behave more like a quasiparticle.}
\end{figure}

Having presented the mean-field Bose-Hubbard and Bogoliubov Hamiltonians for the condensate, we now introduce an additional internal excited state $r^\dag \ket{\text{vac}}$ and a bare bosonic state $a_{j_0}^\dag \ket{\text{vac}}$ at the lattice site $j=j_0$. The bosons at this site experience the laser-induced transition with a coupling strength of $\Omega$, and the dissipative action resulted from measuring  bosons in the internal state is driven by a jump operator $\Gamma$. We shall consider two cases of lossy dynamics: \textit{particle loss} represented by $\Gamma=\sqrt{\kappa}r$, and \textit{particle counting (lossless)} represented by $\Gamma=\sqrt{\kappa}r^\dagger r$, both of which are assigned with a damping rate $\kappa$. Overall, we consider a total system with the Hamiltonian and jump operator
\begin{align}
    H &=H_0+V=H_{MF}+\Delta r^\dagger r+V&\label{BH-H}\\
    \Gamma &=\sqrt{\kappa}r, \quad\text{or}\quad\Gamma = \sqrt{\kappa}r^\dagger r&\label{jump}.
\end{align}
where the mean-field Hamiltonian $H_{MF}$ in the bare particle and quasiparticle representations take the form,
\begin{align}
        H_{MF} &= -t\sum_{j}(a_j^\dagger a_{j+1}+a^\dagger_{j+1}a_j)\nonumber\\
        &+(2Un_0-\mu)\sum_{j}a_j^\dagger a_j\nonumber\\
        &+Un_0\sum_j(a_j^2+a^{\dagger2}_j)+E_{0}\\
        &=\sum_{k\neq0}E_{b,k}b^\dagger_kb_k+E_{\text{cond}}.
\end{align}
Similarly, in these two representations, the interaction term $V$ between the condensate and bosons in the probe state $\ket{r}$ is
\begin{align}
    V&=\Omega \sqrt{n_0}(r+r^\dagger)+\Omega(r^\dagger a_{j_0}+a_{j_0}^\dagger r)\\
        &=\Omega \sqrt{n_0}(r+r^\dagger)
        +\frac{1}{\sqrt{N}}\Omega\sum_{k\neq0}[r^\dagger e^{-ilk}(u_kb_k-v_kb^\dagger_{-k})\nonumber\\
        &+(u_kb_k^\dagger-v_kb_{-k}) e^{ilk}r].&\label{V_single_site}
\end{align}
Note that, in the interaction term $V$ given by Eq.~(\ref{V_single_site}), the bosonic operators $a_{j}$ contains only the non-zero $k$ modes $a_{j}=\frac{1}{\sqrt{N}}\sum_{k\neq0}a_ke^{-ijk}$, as mentioned in the derivation for the mean-field Hamiltonian in Eq.~(\ref{bose_hubbar_bare}). We also express this coupling term in the Bogoliubov basis to later facilitate the narrow measurement bandwidth discussion.

\subsection{Wide Measurement Bandwidth}
In the case with a wide measurement bandwidth, we similarly assume a large detuning $\Delta$ and large damping rate $\kappa$ compared to the drive $\Omega$. In terms of adiabatic elimination of the fast-evolving $r$-species that is being measured, we are only concerned with the linear Adjoint Master equations of the modes $a_{j_0}$ and $r$, in order to trace out $r$. Overall in this regime, after adiabatic elimination, the effective 1D ring Hamiltonian and jump operators are given by. In this regime, we find that,
\begin{align}
    H_{\text{eff,wide}} &= H_{MF}-\frac{\Omega^2\Delta}{\Delta^2+\kappa^2/4}\left(a_{j_0}^\dagger+\sqrt{n_0}\right)\left(a_{j_0}+\sqrt{n_0}\right)\\
    \Gamma_{\text{eff,wide}}&=\sqrt{\frac{\Omega^2}{\Delta^2+\kappa^2/4}\kappa}(a_{j_0}+\sqrt{n_0})\\
    &\text{if $\Gamma=\sqrt{\kappa}r$}\nonumber\\
    \Gamma_{\text{eff,wide}}&=\sqrt{\frac{\Omega^2}{\Delta^2+\kappa^2/4}\kappa}(a_{j_0}^\dagger+\sqrt{n_0})(a_{j_0}+\sqrt{n_0})\label{count}\\
    &\text{if $\Gamma=\sqrt{\kappa}r^\dagger r$}.\nonumber
\end{align}
In deriving the effective Lindblad form and associated jump operators for $a_{j_0}$, we should emphasize that, as far as the Adjoint Master equation is concerned, the dissipator of $\mathcal{L}^\dagger a_{j_0}$ after adiabatically eliminating $r$ is the same, i.e $\mathcal{L}^\dagger[\Gamma_{\text{eff,wide}}](a_{j_0})=-\frac{\kappa}{2}\frac{\Omega^2}{\Delta^2+\frac{\kappa^2}{4}} a_{j_0}$ whether $\Gamma\propto r$ or $\Gamma\propto r^\dagger r$. Nonetheless, a rigorous way to  determine the exact form of $\Gamma_{\text{eff,wide}}$ is to derive the stochastic Heisenberg equations for $a_{j_0}$ and $r$, which corresponds to Eq.(6) supplemented with noise terms. The stochastic equations reveal that the noise operator resulting from a linear jump $\Gamma=\sqrt{\kappa}r$ differs fundamentally from that of a quadratic jump $\Gamma=\sqrt{\kappa}r^\dagger r$. By performing adiabatic elimination in the framework of stochastic equations,  we can obtain the appropriate noise terms that attribute $\Gamma_{\text{eff,wide}}$ to either $a_{j_0}$ or $a_{j_0}^\dagger a_{j_0}$.

Since we previously define $n_0=\sqrt{N_0/N}$ and $a_{j_0}=\sqrt{1/N}\sum_{k\neq0}a_ke^{-ijk}$, the displaced field $a_{j_0}+\sqrt{n_0}=\sqrt{1/N}\sum_{k}a_ke^{-ijk}$ is nothing but the original bosonic operator $a_{j_0}$. Therefore, regarding the effective system and effective measurement, unsurprisingly, similar to the bare state of the two-level atom being effectively measured under wide bandwidth, here we also obtain an effective measurement on the bare bosons at the site $l$, as indicated by the jump operators after adiabatic elimination. The procedure also yields a Stark-like shift of $\frac{\Omega^2\Delta}{\Delta^2+\kappa^2/4}$ in the chemical potential at that site, in place of the coupling.

\subsection{Narrow Measurement Bandwidth}

In the narrow measurement bandwidth, we only assume that the coupling strength $\Omega$ to be so weak that the interaction $V$ enters perturbatively into the condensate dynamics, with the SW transformation generator $S$ given by,
{\small
\begin{align}\label{BH-SW}
    S &= \frac{1}{\sqrt{N}}\sum_{k\neq0}e^{-ikj_0}\left(\frac{\Omega}{\Delta-E_{b,k}}u_kr^\dagger b_k-\frac{\Omega}{\Delta+E_{b,k}}v_kr^\dagger b_{-k}^\dagger\right)\nonumber\\
    &\quad-\sqrt{n_0}\frac{\Omega}{\Delta}r-\text{h.c}.
\end{align}
}%
In a similar manner to the two-level atom, the transformation $e^{-S}$ dresses the Bogoliubov energies, and shifts the detuning $\Delta$ of the $r$ bosons. Unique to the case of the Bose-Hubbard condensate, however, it also induces long-range interactions, and introduces squeezing between the Bogoliubov quasiparticles at different momentum modes $k$ and the $r$ bosons, with squeezing strengths of order $\Omega^2$, and displaces the condensate the site $j=j_0$. In the case of particle loss $\Gamma = \sqrt{\kappa}r$, after performing adiabatic elimination of the external $r$ bosons around an energy pole $\epsilon$ (detailed derivation in App.~\ref{appendix:b}), we find the effective Hamiltonian to be,
\begin{align}
    &H_{\text{eff,narrow}}=H_{MF}-\sum_{k,k'\neq0}T_{k,k'}(b_k^\dagger b_{k'}+b_{-k}^\dagger b_{-k'})\nonumber\\
        &-\sum_{k,k'\neq0}\Omega_{k}\left[b_{k}b_{-k'}e^{-ij_0(k-k')}+b^\dagger_kb^\dagger_{-k'}e^{ij_0(k-k')}\right]\nonumber\\
        &+\frac{\sqrt{n_0}}{2\sqrt{N}}\frac{\Omega^2}{\Delta}\sum_{k\neq0}\left[e^{-ikj_0}(u_kb_k-v_kb_{-k}^\dagger)+\text{h.c}\right]\nonumber\\
        &-\frac{(\Delta'+\epsilon)\kappa^2/4}{(\Delta'+\epsilon)^2-4\gamma_r^2+\kappa^2/4}B^\dagger B&\label{Heff_loss},
\end{align}
and the effective jump operator is,
\begin{align}\label{destroy}
\Gamma_{\text{eff,narrow}} &=\sqrt{\kappa}\left(1+\frac{\kappa^2/4}{(\Delta'+\epsilon)^2-4\gamma_r^2+\kappa^2/4}\right)^{\frac{1}{2}}B.
\end{align}
Here, for mathematical convenience, we have defined the quasiparticle superposition operator $B$ as 
\begin{align}
    B&=\sum_{k\neq0}\frac{e^{-ikj_0}}{\sqrt{N}}\left(\frac{\Omega}{\Delta-E_{b,k}}u_kb_k-\frac{\Omega}{\Delta+E_{b,k}}v_kb^\dagger_{-k}\right)\nonumber\\
       &-\frac{\Omega}{\Delta}\sqrt{n_0},
\end{align}
which represents the general detuning-dependent phonon mode being measured. 

In the dressed picture, $\gamma_r\equiv-\frac{1}{N}\sum_{k\neq0}\frac{E_{b,k}\Omega^2}{\Delta^2-E_{b,k}^2}$ is the squeezing strength of the $r$ bosonic mode, while $\Delta'\equiv\Delta+\frac{1}{N}\left(\sum_{k\neq0}\frac{\Omega^2(\Delta-A_k)}{\Delta^2-E_{b,k}^2}\right)$ is the dressed detuning. These dressed effects on the $r$ bosons are absorbed into the last two terms of $H_{\text{eff,narrow}}$, which results from adiabatically eliminating these probe bosons. Among the Bogoliubov modes, the transformation $e^{-S}$ introduces long-range interactions, given by strength $T_{k,k'}\equiv\frac{1}{N}\left(\frac{\Omega^2}{\Delta-E_{b,k}}u_ku_{k'}-\frac{\Omega^2}{\Delta+E_{b,k}}v_kv_{k'}\right)\cos(ak)$, effective squeezing $\Omega_{k}\equiv\frac{1}{N}\frac{ E_{b,k}\Omega^2}{\Delta^2-E_{b,k}^2}$. From the form in Eq.~(\ref{destroy}), we can see that the effective jump operator, generally, corresponds to measurement of a displaced superposition of Bogoliubov quasiparticles. 

The effective Hamiltonian in the case of particle counting $\Gamma=\sqrt{\kappa}r^\dagger r$, after adiabatically eliminating $r$, has a very similar structure, which implies the same physics, albeit with slightly different coefficients after tracing out the probe bosonic mode $r$. When there is no particle loss, the narrow bandwidth effective Hamiltonian is given by,
\begin{align}
        &H_{\text{eff,narrow}}=H_{MF}-\sum_{k,k'\neq0}T_{k,k'}(b_k^\dagger b_{k'}+b_{-k}^\dagger b_{-k'})\nonumber\\
        &-\sum_{k,k'\neq0}\Omega_{k}\left[b_{k}b_{-k'}e^{-ij_0(k-k')}+b^\dagger_kb^\dagger_{-k'}e^{ij_0(k-k')}\right]\nonumber\\
        &+\frac{\sqrt{n_0}}{2\sqrt{N}}\frac{\Omega^2}{\Delta}\sum_{k\neq0}\left[e^{-ikj_0}(u_kb_k-v_kb_{-k}^\dagger)+\text{h.c}\right]\nonumber\\
        &-\frac{(\Delta'+\epsilon)\kappa^2/4}{(\Delta'+\epsilon)^2-4\gamma_r^2+\bar{\kappa}^2/4}B^\dagger B&\label{Heff_lossless},
\end{align}
and the effective jump is, 
\begin{align}\label{count}
    \Gamma_{\text{eff,narrow}}&=\sqrt{\frac{\kappa}{\Pi}}\left(1+\frac{\kappa\bar{\kappa}/4}{(\Delta'+\epsilon)^2-4\gamma_r^2+\bar{\kappa}^2/4}\right)^{\frac{1}{2}}\nonumber\\
    &\times B^\dagger B
\end{align}
where we define $\bar{\kappa}\equiv\kappa(1+\Pi)$ for convenience, with $\Pi \equiv [B,B^\dagger]=\frac{1}{N}\sum_{k\neq0}\left[\frac{\Omega^2}{(\Delta-E_{b,k})^2}u_k^2-\frac{\Omega^2}{(\Delta-E_{b,k})^2}v_k^2\right]$ being the commutator of the probe superposition $B$.
Given the structure of the superposition $B$ over all non-zero momenta, we can attempt to measure a phonon mode of momentum norm $|k|=|q|\neq0$ by tuning $\Delta$ to be near-resonant with the corresponding Bogoliubov energy $E_{b,q}$. This is achievable provided that the SW constraint $\Omega<<|\Delta-E_{b,k}|\forall k\neq0$ remains valid. Unfortunately, fine-tuning $\Delta$ alone cannot fully resolve Bogoliubov quasiparticles propagating with momentum $q$ (i.e reducing the effective jump operator to simply $\Gamma_{\text{eff,narrow}}\propto b_{q}$). This is because the single-site drive is agnostic of directions in momentum space, exerting "kicks" in all the direction. 

One way to circumvent this problem is to laser drive \textit{all sites} to the probe internal state $\ket{r}$, and modulate the laser strength $\Omega$ with a wavevector $p$. Mathematically, we first re-define the interaction term $V$ as followed,
\begin{equation}\label{Vall}
    V_{\text{all sites}} = \Omega\sum_{j}\left(r_j^\dagger a_j e^{ipj}+a_j^\dagger r_je^{-ipj}\right),
\end{equation}
and allow the bare particles in the $\ket{r}$ to tunnel freely, while still assuming a dominant condensed population of the $\ket{a}$ particles. This physical process is realizable with Bragg spectroscopy, where the wavevector $p$ originates from the momentum kick of two laser beams. In $V_{\text{all sites}}$, the spatial phase modulation serves to couple every condensate quasiparticle of momentum $k$ with an uncondensed bare particle of momentum $k-p$ in the probe state. Consequently, the SW generator $S_{\text{all sites}}$ that decouples the condensate from the probe bosons takes the form,
\begin{align}\label{Sall}
    S_{\text{all sites}} &=\sum_{k\neq 0}r^\dagger_{k-p}\left(\frac{\Omega u_k}{\mathcal{E}_{k-p}-E_{b,k}}b_k-\frac{\Omega v_k}{\mathcal{E}_{k-p}+E_{b,k}}b_{-k}^\dagger\right)\nonumber\\
    &- \frac{\Omega\sqrt{n_0}}{\mathcal{E}_{-p}}r_{-p}-\text{h.c},
\end{align}
where $\mathcal{E}_k \equiv \Delta-2t\cos k$ is the free particle dispersion of the bare particles in the uncondensed probe state. We observe that the momentum difference $p$ in the denominators of the decoupling terms, in the SW generator $S_\text{all sites}$, gives rise to \textit{anisotropic resonances} between the $\pm k$ modes. Within the validity regime of the SW transformation, defined by $\Omega\ll |\mathcal{E}_{k-p}-E_{b,k}|\forall k$, this anisotropy is embedded into the near-resonant values of $\Delta_{\pm q}$ between any modes of equal and opposite momenta. Specifically, a finite gap of $|4t\sin q\sin p|$ emerges, allowing us to distinguish the modes in our earlier protocol. 

Outside the regime - particularly near perfect resonance of the measure mode $q$, where $\mathcal{E}_{q-p}-E_{b,q}$, the SW approximation breaks down and our protocol no longer applies. Instead, we expect a dominant \textit{direct} interaction between the condensed quasiparticles $\ket{b_q}$ and the probe particles $\ket{r_{q-p}}$. Nonetheless, this interaction can still be exploited for a mode-selective measurement protocol of $q$, via probing the corresponding detuned particle state.

We shall leave the full analytical expressions for the relevant damping rates, shown below, in App.~\ref{appendix:c}. In both cases of particle loss $\Gamma=\sqrt{\kappa}r$ and particle counting (lossless), the effective jump operators (in $k$-space, $k\neq0$) are given by
\begin{align}
B_{k} &=-\frac{\Omega u_k}{\mathcal{E}_{k-p}-E_{b,k}}b_k+\frac{\Omega v_k}{\mathcal{E}_{k-p}+E_{b,k}}b_{-k}^\dagger\\
\Gamma_{k}&=\sqrt{\kappa}\sqrt{1+\tilde{\kappa}_1(k)}B_{k}\nonumber\\
&\quad\text{if }\Gamma_j=\sqrt{\kappa}r_j\label{all_loss}\\
\Gamma_{k}&=\sqrt{\kappa}\sqrt{\frac{1}{\Pi_{\text{all sites},k}}+\frac{\tilde{\kappa}_2(k)}{\Pi_{\text{all sites},k}}}B_{k}^\dagger B_k\nonumber\\
&\quad\text{if }\Gamma_j=\sqrt{\kappa}r_j^\dagger r_j\label{all_lossless},
\end{align}
where $\Pi_{\text{all site}}(k)\equiv[B_{k},B_{k}^\dagger]=\frac{\Omega^2}{(\mathcal{E}_{k-p}-E_{b,k})^2}u_k^2-\frac{\Omega^2}{(\mathcal{E}_{k-p}+E_{b,k})^2}v_k^2$ and $\tilde{k}_1(k),\tilde{k}_2(k)$ are dimensionless coefficient adjustments similar to those in the case of single-site particle counting. Note that the (unnormalized) operator $B_k$ has a well-defined inverse-Fourier transform in real space, which itself arises from the SW correction $[S,\Gamma_j]$ associated with each local jump operator $\Gamma_j$. Because the effective damping rate coefficients $\tilde{k}_1(k)$ and $\tilde{k}_2(k)$ depend on $k$, the resulting real-space effective jump operator $\Gamma_{\text{eff}-j}$ is expected to contain the original local jump term $\Gamma_j$ at leading order, together with a convolution correction weighted by the inverse-Fourier transforms of $\tilde{k}_1(k),\tilde{k}_2(k)$.

\begin{figure}[ht]
\centering
\begin{minipage}{\columnwidth}
  \centering
  \includegraphics[width=\linewidth]{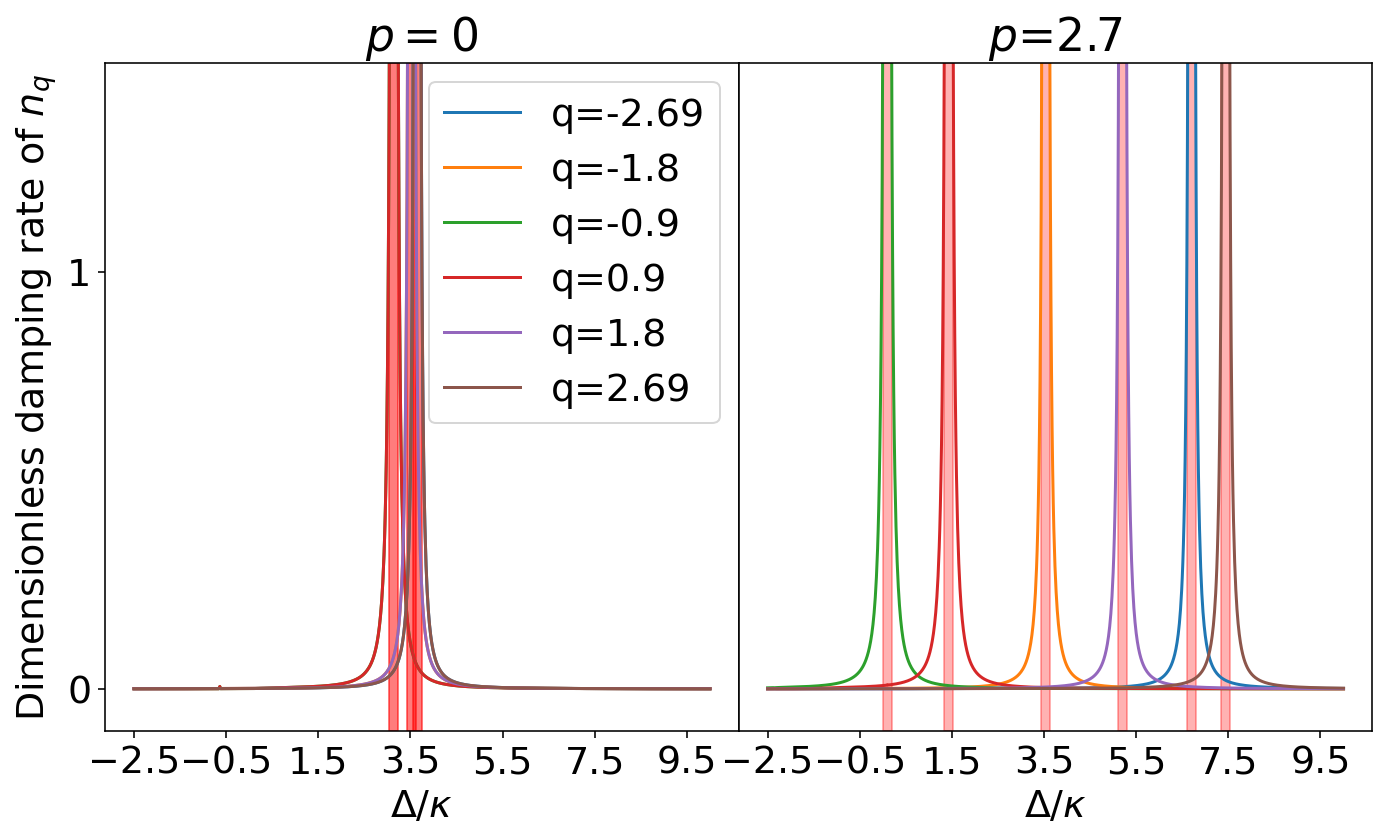}
  \caption*{(a) Effective particle loss rate ($\Gamma_{\text{eff}}\propto b_q)$ }
\end{minipage}

\begin{minipage}{\columnwidth}
  \centering
  \includegraphics[width=\linewidth]{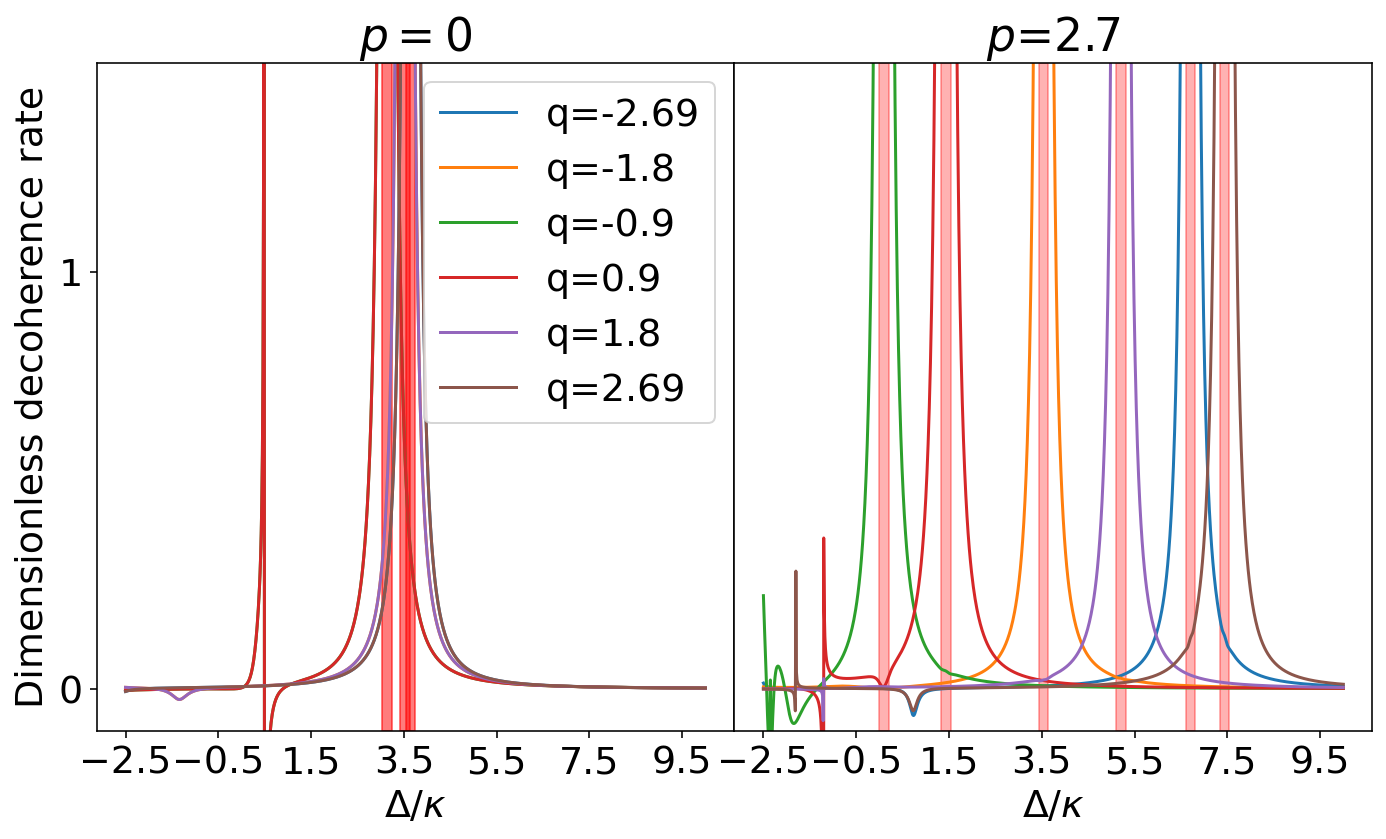}
  \caption*{(b) Effective particle dephasing rate $(\Gamma_{\text{eff}}\propto b_q^\dagger b_q)$}
\end{minipage}
\caption{Damping rate resonances. The shaded red regions are forbidden by the SW condition.}
\label{damping_rate}
\end{figure}

Nonetheless, for the purpose of our protocol, we are primarily interested in the self-damping rate of the desired Bogoliubov mode $q$ when the detuning $\Delta$ is tuned near resonance, thus the results in Eq.~(\ref{all_loss}) and Eq.~(\ref{all_lossless}) suffice. Recall that near resonance with the measurement mode $b_q$, the effective jump operators can approximately be reduced to either $b_q$ or $b_q^\dagger b_q$, depending on the underlying loss mechanism. Using these expressions, in Fig.~\ref{damping_rate}, we plot (a) the effective quasiparticle population decay rate and (b) the effective quasiparticle dephasing rate, for a condensed system of six modes, as a function of $\Delta$. To highlight the anisotropy,  we compare these quantities at $q=0$ and $q\neq0$.

In both damping mechanisms, for $p=0$ (i.e., without phase modulation), the damping rates are completely isotropic, and the peaks are tightly spaced. This is because the resonance value per $|q|$ mode, set by the energy difference $\Delta_q = 2t\cos(q) - E_{b,q}$ (for $p=0$), is relatively small. When a nonzero phase modulation $p$ is introduced, anisotropy emerges and the resonance peaks become more widely separated. This separation is advantageous, since we cannot access perfect resonance without violating the SW perturbation conditions, and therefore measurement can only be performed in its vicinity. 

Although not relevant to our measurement protocol, in the case of dephasing, poles also appear when $\Pi_{\text{all sites},k}=0$. At these values, defined by $\Delta_{\Pi}(k) = 2 t\cos(k-p)-\frac{u_k-v_k}{u_k+v_k}$ or $\Delta_{\Pi}(k) = 2 t\cos(k-p)-\frac{u_k+v_k}{u_k-v_k}$, the (unnormalized) operator $B_k$ simply behaves classically.

\section{\label{sec:level5}Measurement-induced Condensate Heating}
In this section, we provide the theoretical evidence for the increase in Bogoliubov quasiparticles over time- interpreted as \textit{quasiparticle heating} - at zero temperature, induced by the weak measurement procedures introduced in Sec.~\ref{sec:level4}, when the system is initial prepared in the condensate ground-state. We shall analyze this heating of the Bose-Hubbard condensate in the two previously discussed measurement bandwidths. We find evidence of quasiparticle increase in the wide measurement bandwidth, with the heating behavior being similar whether or not particle loss occurs. On the other hand, under our narrow measurement bandwidth protocol described, although heating persists across most Bogoliubov modes, the selective near-resonant tuning of $\Delta$ actually suppresses the heating of the desired probe mode. Therefore, this heating behavior showcases the protocol's ability to detect quasiparticle in the narrow measurement bandwidth. 

\subsection{Quasiparticle Heating in the Wide Measurement Bandwidth:}

In the effective mean-field Hamiltonian after adiabatically eliminating the fast laser drive, we recall that the site $j=j_0$ now has a dressed chemical potential, and the effective weak measurement procedure is akin to removing a boson at the same site. We can write the aperiodic Bose-Hubbard Hamiltonian due to particle removal at a site as,
\begin{align}\label{BoseHubbard}
        H_{BH,AP} &= -t\sum_{\text{n.n}}a^\dagger_j a_{j'} +\frac{U}{2}\sum_{j}a^\dagger_ja^\dagger_ja_ja_j\nonumber\\
        &-\sum_j(\mu+\epsilon_j)a^\dagger_j a_j,
\end{align}
where we define the on-site correction to the chemical potential at $j=j_0$ as $\epsilon_i=\delta_{j,j_0}\frac{\Omega^2\Delta}{\Delta^2+\kappa^2/4}$. According to ~\cite{wang_bogoliubov_2016}, in the condensation limit, an aperiodic change in the chemical potential simply introduces a squeezing term in k-space,
\begin{align}\label{aperiodicH}
        H_{MF,AP} &= E_{\text{cond}}+\sum_{k\neq0}\left[ a_k^\dagger a_k+\frac{Un_0}{2}(a_ka_{-k}+a_k^\dagger a_{-k}^\dagger)\right]\nonumber\\
        &\quad-\sqrt{n_0}\sum_{k\neq0}\left(\epsilon_k a_k^\dagger+\epsilon_{-k}a_{k}\right),
\end{align}
where the Fourier transform of $\epsilon_j$ is given by
\begin{equation}
    \epsilon_k =\frac{1}{\sqrt{N}}\sum_{j}e^{ijk}\epsilon_j=\frac{1}{\sqrt{N}}e^{ikl}\frac{\Omega^2\Delta}{\Delta^2+\kappa^2/4}.
\end{equation}
The Hamiltonian in Eq.~(\ref{aperiodicH}) can still be diagonalized by a Bogoliubov transformation, displaced by the chemical potential correction $\epsilon_k$. For $k\neq0$, the Bogoliubov transformation \cite{wang_bogoliubov_2016} is given by,
\begin{align}
        a_k &= u_kb_k-v_kb_{-k}^\dagger+\frac{\sqrt{n_0}}{A_k+Un_0}\epsilon_k\\
        a^\dagger_k &= u_kb^\dagger_k-v_kb_{-k}+\frac{\sqrt{n_0}}{A_k+Un_0}\epsilon_{-k}.
\end{align}
And the Bogoliubov Hamiltonian therefore contains a correction to the vacuum energy due to $\epsilon_k$,
\begin{align}\label{aperiodicdiag}
        H_{MF,AP} &=\sum_{k\neq0}E_{b,k}b^\dagger_kb_k+E_{\text{cond}}+\sum_{k\neq0}\frac{n_0}{A_k+Un_0}|\epsilon_k|^2.
\end{align}
Recall that the effective measurement procedure removes the bare bosons at site $j=l$, and the effective damping rate is $\kappa_{e}\equiv\kappa_{\text{eff,wide}}=\frac{\Omega^2}{\Delta^2+\kappa^2/4}\kappa$. Let us first consider the particle loss case, where the effective jump operator is given by $\Gamma_{\text{eff,wide}}=\sqrt{\kappa_e}a_{j_0}$. Under the displaced Bogoliubov transformation, the jump operator takes the form
\begin{align}
    \Gamma &= \sqrt{\kappa_e}a_{j_0}\nonumber\\
    a_{j_0} &= \sqrt{n_0}+\frac{1}{\sqrt{N}}\sum_{k\neq0}e^{-ikl}a_k\nonumber\\
    &=\sqrt{n_0}+\frac{1}{\sqrt{N}}\sum_{k\neq0}e^{-ikl}\left(u_kb_k-v_kb_{-k}^\dagger\right)\nonumber\\
    &\quad+\frac{1}{\sqrt{N}}\sum_{k\neq0}e^{-ikl}\frac{\sqrt{n_0}}{A_k+Un_0}\epsilon_k.
\end{align}
After establishing the mean-field Hamiltonian and jump operator under the new Bogoliubov transformation, in order to investigate the condensate heating over time, we simply consider the Adjoint master equation for the number of Bogoliubov quasiparticles per $k$ mode, as given by
\begin{align}
    \expval{b_k^\dagger b_k}&=\text{Tr}(\mathcal{L}^\dagger(b_k^\dagger b_k)\rho)&\label{bdb}\\
    \mathcal{L}^\dagger (b^\dagger_k b_k)&=\frac{1}{2}\left(2\Gamma^\dagger b^\dagger_k b_k\Gamma -\{\Gamma^\dagger \Gamma,b^\dagger_k b_k\}\right).
\end{align}

\begin{figure}[h]
\includegraphics[width=8cm]{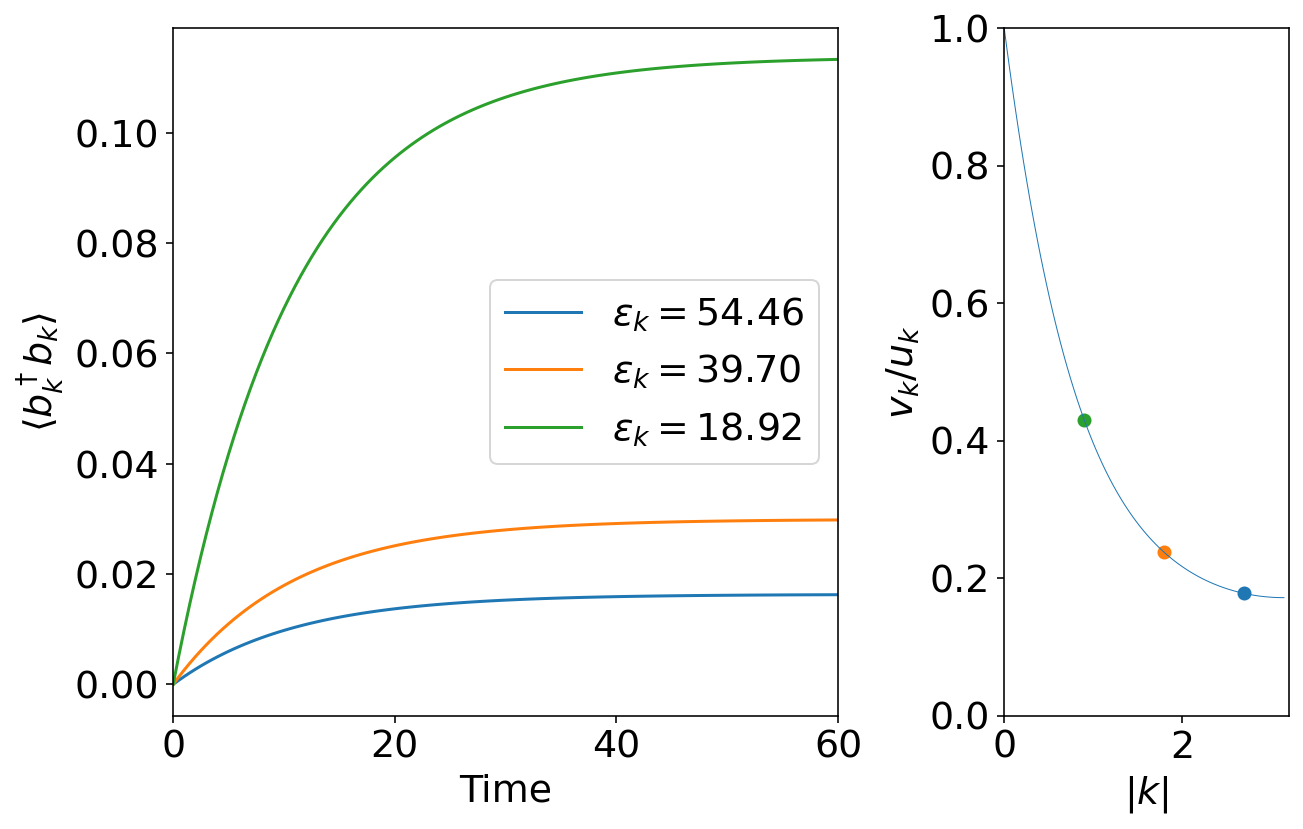}
\caption{\label{fig:heating_removal} Quasiparticle heating over time for $N=7$ due to loss.}
\end{figure}

The right plot of Fig. \ref{fig:heating_removal} shows the numerical time evolution of $\expval{b_k^\dagger b_k}$ on a 7-site periodic lattice when dissipation occurs during measurement, whereas the left plot displays the respective Bogoliubov energies on the $u_k/v_k$ curve, as seen in Fig. \ref{fig:bogoliubov}. Due to the system's isotropicity, we are only concerned with modes of distinct momentum norms $|k|$. After the system is prepared in the ground-state of the mean-field Bose-Hubbard Hamiltonian in Eq.~(\ref{aperiodicdiag}), at which no Bogoliubov quasiparticles are present in the condensate, we can see that the weak measurement induces quasiparticle increase that scales inversely the Bogoliubov energy. Since the jump operator is a linear combination of quasiparticle creation and annihilation, we can see that the quasiparticle occupation number of each mode increases towards a stable gain value. This gain in quasiparticle is proportional to the Bogoliubov coefficient $v_k^2$ per mode $k$, which appears as a "vacuum" gain in the Master equations of all relevant quadratic variables $\expval{b_{k}^\dagger b_{k'}}$.

In addition, the rate of heating up to this steady value is proportional to the damping rate $\kappa_e/N$ of the effective weak measurement procedure, where $\kappa_e$ is previously given by
\begin{equation}
    \kappa_e=\frac{\Omega^2}{\Delta^2+\kappa^2/4}\kappa
\end{equation}
This expression shows that the heating rate decreases with increasing detuning $\Delta$ of the external bosons in probed in the original experiment, and can only increase as a function of $\kappa$ up to a critical value. Beyond this value, excessive damping in observing the $r$ bosons inadvertently dampens out the heating.

\begin{figure}[h]
\includegraphics[width=8cm]{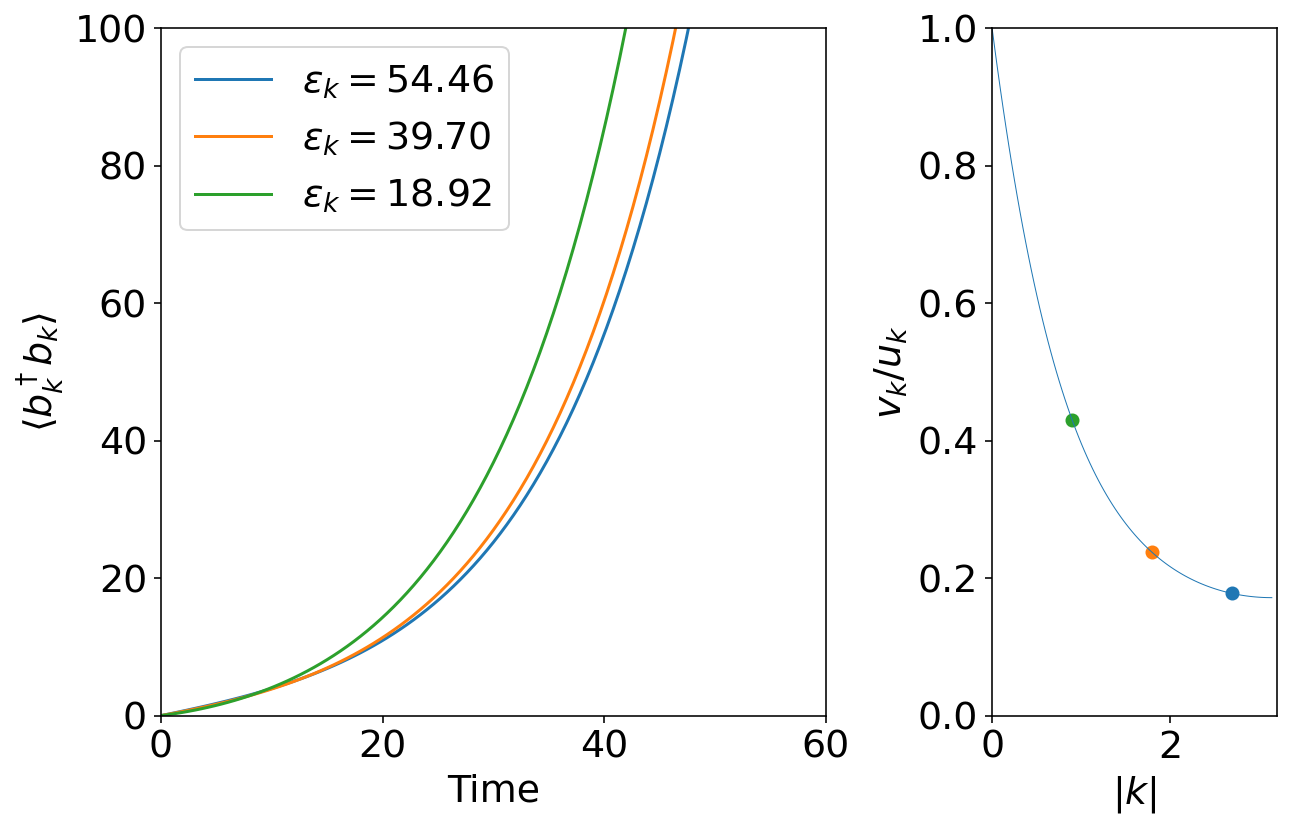}
\caption{\label{fig:heating_counting} Quasiparticle heating over time for $N=7$ without loss.}
\end{figure}

On the other hand, when dissipation does not occur, or equivalently $\Gamma = \sqrt{\kappa_e}a_{j_0}^\dagger a_{j_0}$, Fig. \ref{fig:heating_counting} shows that the quasiparticles instead heat up indefinitely. The exponential-like rate of heating, per mode $k$, remains proportional to the effective damping rate $\kappa_e$, and the amount of quasiparticle heating gain also scales with $v_k^2$.

\subsection{Heating in the Narrow Measurement-Bandwidth Regime} 
In the narrow measurement-bandwidth regime, we recall that the condensate evolves under a different effective Hamiltonian (see App.~\ref{appendix:c}), and undergoes the weak measurement induced by the jump operators in Eq.~(\ref{all_loss}) and Eq.~(\ref{all_lossless}). Because of the coupling potential in Eq.~(\ref{Vall}), in this regime, the effective jump operators only couples quasiparticles modes with momenta $\pm k$, such that dissipation induces quasiparticle exchange channels exclusively between modes of equal momentum magnitude. Under our protocol for quasiparticle detection, the introduction of $\textit{anisotropic}$ driving, via the phase modulation with wavevector $p$ in Eq.~(\ref{Vall}), breaks the symmetry and leads to distinct heating dynamics between $\pm k$ modes. Furthermore, by tuning $\Delta$ to be in near-resonance with a selected Bogoliubov mode $q$, quasiparticle heating in the observed mode is selectively suppressed.

\begin{figure}[h]
\includegraphics[width=8cm]{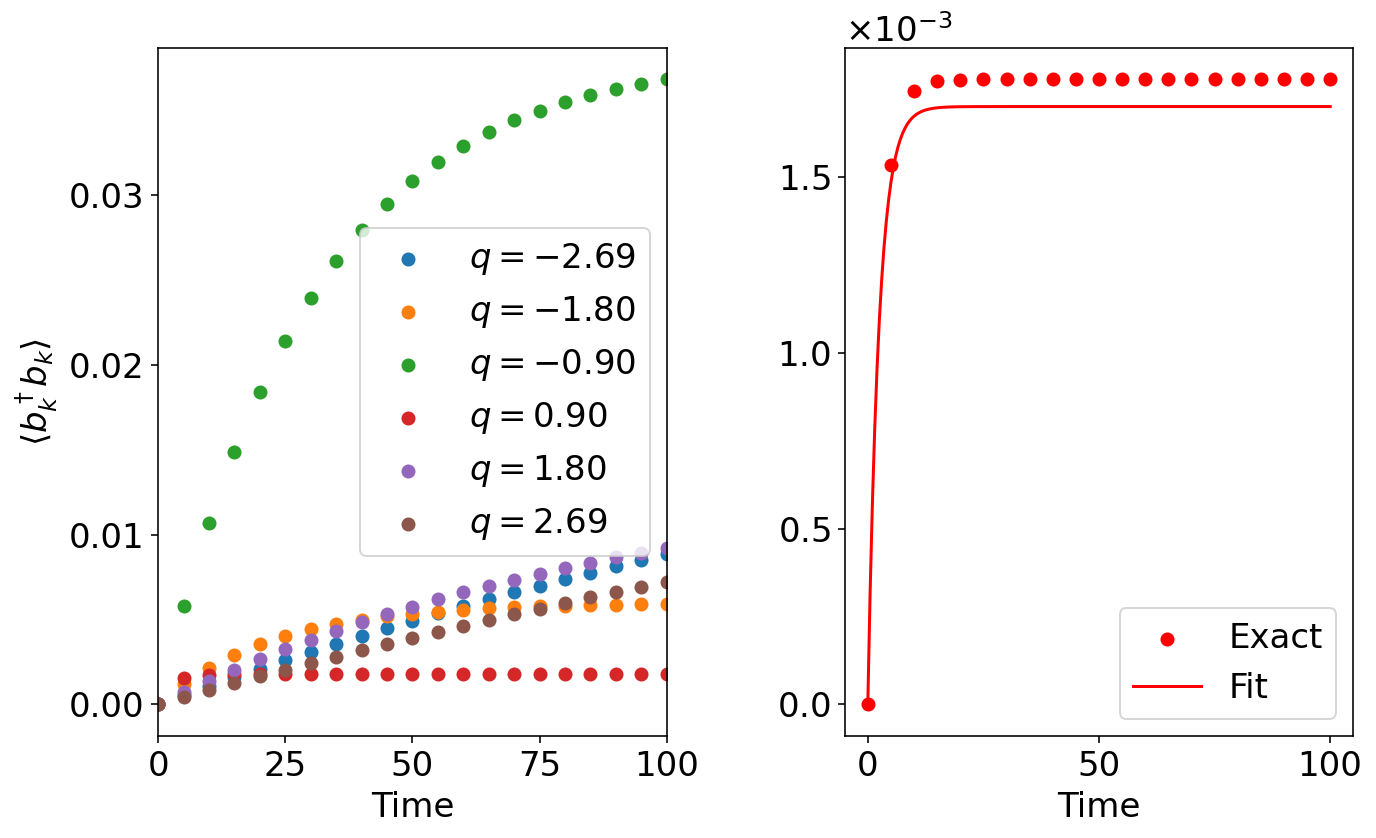}
\caption{\label{fig:narrow_removal_heating} Quasiparticle heating of a 7-site system over time due to loss, in the narrow measurement bandwidth. In both figures, we tune $\Delta$ to be near $\Delta_q = 2t\cos(q-p)-E_{b,q}$ for $q=0.9$, such that the quasiparticle heating of mode $q=0.9$ (red) is suppressed. The left figure plots numerical average quasiparticle increase in all 6 quasiparticle modes, whereas the right figure compares the numerical result against the asymptotic fit given in Eq.~(\ref{fit}). } 
\end{figure}

In Fig.~\ref{fig:narrow_removal_heating}, we plot the numerical time evolution of the quasiparticle occupation $\expval{b_k^\dagger b_k}$ on a 7-site lattice, when dissipation occurs under the effective jump operator in Eq.~(\ref{all_loss}). With the same wavevector modulation in Fig.~\ref{damping_rate}(a), we select the observed mode to be $q=0.9$, by setting $\Delta$ to be close to its resonance condition,
\begin{align}
    \Delta_q &= 2t\cos(q-p)-E_{b,q}
\end{align}
,where $E_{b,q}$ is the corresponding Bogoliubov energy. Consequently, while the remaining quasiparticle modes exhibit heating dynamics similar to that found in the wide bandwidth regime, the heating gain of the observed mode $q=2$ is strongly suppressed. This suppression can be shown by expanding the dissipator acting on $\expval{b_q^\dagger b_q}$ in the limit $\Delta\to\Delta_q$,
\begin{align}
    \mathcal{L}_{\text{diss}}^\dagger(b_q^\dagger b_q)&\approx-\kappa(1+\tilde{\kappa}_1(q))\frac{\Omega^2u_q^2}{(\mathcal{E}_{q-p}-E_{b,q})^2}b_q^\dagger b_q\nonumber\\
    &+\kappa(1+\tilde{\kappa}_1(-q))\frac{\Omega^2v_q^2}{(\mathcal{E}_{q+p}+E_{b,q})^2}.
\end{align}
which solves for,
\begin{align}\label{fit}
    \expval{b_q^\dagger b_q}(t) &\approx\frac{1+\tilde{\kappa}_1(-q)}{1+\tilde{\kappa}_1(q)}\frac{v_q^2(\mathcal{E}_{q-p}-E_{b,q})^2}{u_q^2(\mathcal{E}_{q+p}+E_{b,q})^2}\nonumber\\
    &\cdot\left[1-\exp\left(-\kappa(1+\tilde{\kappa}_1(q))\frac{\Omega^2u_q^2}{(\mathcal{E}_{q-p}-E_{b,q})^2}t\right)\right].
\end{align}
As $\Delta$ approaches $\Delta_0$, the steady-state heating gain $\frac{1+\tilde{\kappa}_1(q)}{1+\tilde{\kappa}_1(-q)}\frac{u_q^2(\mathcal{E}_{q+p}+E_{b,q})^2}{v_q^2(\mathcal{E}_{q-p}-E_{b,q})^2}$ becomes vanishingly small. To arrive at the asymptotic form in Eq.~(\ref{fit}), we neglect the quasiparticle gain that arises from two-mode squeezing between $\pm k$ pairs, which is generated by the effective jump operator in Eq.~(\ref{all_loss}). This contribution accounts for the small gap between the exact numerics and asymptotic fit, shown in the right panel of Fig.~\ref{fig:narrow_removal_heating}. This approximation is justified because, under the near-resonant tuning of our measurement protocol, the dominant contribution is governed by the large self-decay term proportional to $b_k^\dagger b_k$, whose coefficient scales as $(\mathcal{E}{q-p}-E{b,q})^{-2}$ and therefore captures the leading-order behavior.

\begin{figure}[h]
\includegraphics[width=8cm]{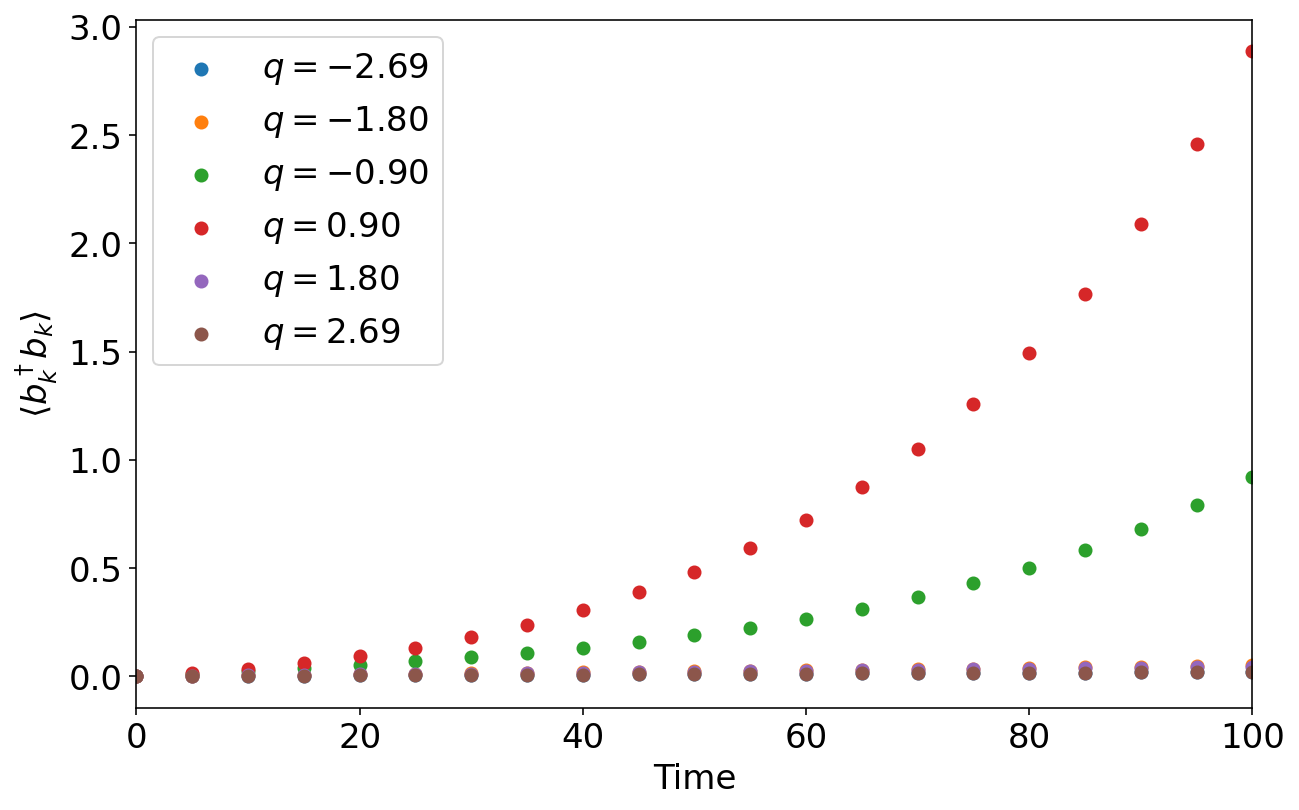}
\caption{\label{fig:narrow_coutning_heating} Quasiparticle heating over time without loss, in the narrow measurement bandwidth. In this plot, we also tune $\Delta$ to be near $\Delta_q = 2t\cos(q-p)-E_{b,q}$ for $q=2$.} 
\end{figure}

In contrast, in the absence of dissipation—when the dynamics are governed by the effective jump operator in Eq.~(\ref{all_lossless})—the quasiparticle gain induced by two-mode squeezing can no longer be neglected, as the large self-decay contribution to $\mathcal{L}^\dagger(b_k^\dagger b_k)$ is absent. Figure~\ref{fig:narrow_coutning_heating} presents the numerical time evolution for the same observed mode at near-resonant detuning $\Delta_q$. In this regime, we observe heating behavior qualitatively similar to the wide-bandwidth lossless case, but with pronounced anisotropy. Notably, unlike the lossy case—where heating of the measured $q=2$ mode is suppressed by the dominant decay channel—the heating here is instead enhanced, since the effect of near-resonant tuning now enters primarily through the quasiparticle gain associated with squeezing.

\section{\label{sec:level6}Outlook}

In this work, we have developed an approach to exploring how measurement probes the quasiparticles of many-body systems in the explicit scenario of cold atoms undergoing optical interrogation. However, our approach may also apply in a variety of other more general scenarios. For example, in situations in which there are known or unknown decoherence mechanisms, our results suggest that the timescale for the dynamics of the decoherence can matter in ascertaining which type of Lindblad term emerges in the low energy many-body dynamics. This may be of particular importance for theories that predict decoherence from, e.g., effects in renormalization flow but also in scenarios such as theories of open quantum systems describing new physics phenomena such as extensions to semiclassical gravity.

Furthermore, the concept that the measurement probe's bandwidth determines the effects of the measurement on the system in many-body cases lends itself to explorations of phenomena often considered standard in exploring the impacts of the fluctuations of vacuum states. For example, in the typical pedagogy of the Casimir effect, one method of understanding considers the vacuum fluctuations as physical, e.g., leading to the observed force. Another method would suggest that it is instead one-loop diagram corrections to the renormalized vacuum. Here our approach would inform this dialogue by asking what happens to such effects when the particles of the theory are under observation. More spectacular (and speculative) examples also can be found in the Kondo effect and other energy-scale-free renormalization problems.

More practically, these ideas are immediately implementable in the ever-versatile cold atom platform, and we hope that others will consider the application of this set of ideas and the exploration of these phenomena in future work.

\begin{acknowledgments}
This work is supported by the Heising-Simons Foundation grant 2023-4467 ``Testing the Quantum Coherence of Gravity" and grant 63121 from the John Templeton Foundation. The opinions expressed in this work are those of the authors and do not necessarily reflect the views of the John Templeton Foundation. Y.-X.W.~acknowledges support from a QuICS Hartree Postdoctoral Fellowship. 
\end{acknowledgments}

\pagebreak
\nocite{*}

\bibliography{bibliography}
\appendix
\section{Schrieffer-Wolff Transformation of the Double-Well Atom}
\label{apppendix:a}
Because the Schrieffer-Wolff transformation $e^{-S}$ is unitary, the generator $S$ is anti-Hermitian. To solve for the Schrieffer-Wolff condition $V+[S,H_{\text{atom}}+\Delta\sigma_{ee}]=0$, we simply assume an ansatz of the form $S = a\sigma_{rL}-a^*\sigma_{Lr}+b\sigma_{rR}-b^*\sigma_{Rr}$, which is akin to a beam-splitter action between the two trapped states and the probe state. Under the SW transformation $S$ in Eq.(\ref{2levelSW}), the decoupled Hamiltonian and dressed jump operator are given by
\begin{align}
        H_{\text{new}} &=e^{S}(H_{\text{atom}}+\Delta\sigma_{22})e^{-S}\nonumber\\
        &\approx H_{\text{atom}} +\frac{1}{2}[S,V]  \nonumber\\
        &\approx H_{\text{atom}}-\delta_L'\sigma_{LL}-\frac{\Omega^2 t}{2\Omega_{SW}^2}(\sigma_{LR}+\sigma_{RL})\nonumber\\
        &\quad+\left[\Delta+\frac{(\Delta-\delta_R)\Omega^2}{\Omega_{SW}^2}\right]\sigma_{rr}\\
        \Gamma_{\text{new}}&=e^{S}\Gamma e^{-S}\nonumber\\
        &\approx \Gamma+[S,\Gamma]\nonumber\\
        &\approx \underbrace{\sqrt{\kappa}\sigma_{rr}+\sqrt{\kappa}\frac{(\Delta-\delta_R)\Omega}{\Omega_{SW}^2}\sigma_{rr}}_{\text{does not contribute to the dissipator of $\ket{L},\ket{R}$}}\nonumber\\
        &-\sqrt{\kappa}\frac{(\Delta-\delta_R)\Omega}{\Omega_{SW}^2}\sigma_{LL}-\sqrt{\kappa}\frac{\Omega t}{\Omega_{SW}^2}\sigma_{LR}
\end{align}

Due to the Pauli algebra of a single atom, we can readily extract the effective jump operators and Hamiltonian of the tunneling atoms by ignoring the decoupled $\sigma_{rr}$ term. However, in the many-body Bose-Hubbard system where cross correlation can arise from the general dissipator, the decoupled probe terms have to be treated with care.

\section{Adiabatic Elimination in the Narrow Bandwidth Regime for the BEC}
\label{appendix:b}
Let's consider the case of particle loss, where the jump operator is $\Gamma = \sqrt{\kappa}r$. Under the SW transformation in Eq.(\ref{BH-SW}), the Hamiltonian in Eq.(\ref{BH-H}) and jump operator becomes,
\begin{align}
        H_{\text{new}} &=H_0+\delta\Delta r^\dagger r\nonumber\\
        &-\gamma_r(r^2+(r^\dagger)^2)-\sum_{k,k'\neq0}\eta_{k,k'}(b_k^\dagger b_{k'}+b_{-k}^\dagger b_{-k'})\nonumber\\
        &-\sum_{k,k'\neq0}\gamma_{k,k'}\left[b_{k}b_{-k'}e^{-il(k-k')}+b^\dagger_kb^\dagger_{-k'}e^{il(k-k')}\right]\nonumber\\
        &+\frac{\sqrt{n_0}}{2\sqrt{N}}\frac{\Omega^2}{\Delta}\sum_{k\neq0}e^{-ikl}(u_kb_k-v_kb_{-k}^\dagger)\nonumber\\
        &+\frac{\sqrt{n_0}}{2\sqrt{N}}\frac{\Omega^2}{\Delta}\sum_{k\neq0}e^{ikl}(u_kb_k^\dagger-v_kb_{-k})&\label{Hnew}\\
       \Gamma_{\text{new}} &=\sqrt{\kappa}r\nonumber\\
       &-\frac{\sqrt{\kappa}}{\sqrt{N}}\sum_{k\neq0}\left(\frac{\Omega}{\Delta-E_{b,k}}u_k b_{k}-\frac{\Omega}{\Delta+E_{b,k}}v_kb^\dagger_{-k}\right)\nonumber\\
       &\cdot e^{ikl}+\sqrt{\kappa}\frac{\Omega}{\Delta}\sqrt{n_0}\nonumber\\
       &=\Gamma -\sqrt{\kappa}B&\label{gamma_new},
\end{align}

where we define,
\begin{align}
    \eta_{k,k'}&\equiv\frac{\cos(k)}{N}\frac{\Omega^2}{\Delta-E_{b,k}}u_ku_{k'}\nonumber\\
    &-\frac{\cos(k)}{N}\frac{\Omega^2}{\Delta+E_{b,k}}v_kv_{k'}\\
    \gamma_{k,k'}&\equiv\frac{1}{N}\frac{ E_{b,k}\Omega^2}{\Delta^2-E_{b,k}^2}\\
    \delta\Delta&\equiv\frac{1}{N}\sum_{k\neq0}\frac{\Omega^2(\Delta-A_k)}{\Delta^2-E_{b,k}^2}\\
    \gamma_r&\equiv\frac{1}{N}\sum_{k\neq0}\frac{E_{b,k}\Omega^2}{\Delta^2-E_{b,k^2}}
\end{align}
The next step is to write down the Adjoint Equations (of the form given in Eq.(\ref{eq:qme.adjoint})) for $b_k$ and $r$ given the decoupled Hamiltonian in Eq.(\ref{Hnew}) and the dressed jump operator in Eq.(\ref{gamma_new}). Unlike the double-well atom example where the original leading order term $\Gamma$ in $\Gamma_{\text{new}}$ does not contribute to the dissipator of the atom's dynamical variables, the external bosonic field $r$ contributes non-trivially to the dissipator of the bare momentum field $b_k$. And therefore, in order to trace out $r$ in the Heisenberg equation of motion of $b_k$ and obtain an effective Lindblad form for the ring system, we assume that the the probed $r$ internal state evolves around some dominant energy pole $\epsilon$ and "enslave" it to the Bogoliubov operators $b_k$. The Adjoint Master equations for $b_k$ and $r$ are given by,
\allowdisplaybreaks
\begin{align}
    \mathcal{L}^\dagger(b_k)&=i[H_{\text{new}},b_k]\nonumber\\
    &+\frac{1}{2}\left(2B^\dagger b_kB-\{B^\dagger B,b_k\}\right)\nonumber\\
    &=i[H'_{MF},b_k]+\mathcal{L}^\dagger_{\text{diss}}[B](b_k) \nonumber\\
    &-\frac{\kappa}{2}e^{ikj_0}\left(\frac{\Omega}{\Delta-E_{b,k}}u_kr+\frac{\Omega}{\Delta+E_{b,k}}v_kr^\dagger\right)&\label{bk}\\
    \mathcal{L}^\dagger(r)&=-i\underbrace{(\Delta+\delta\Delta)}_{\Delta'}r-i2\gamma_rr^\dagger-\frac{\kappa}{2}r\nonumber\\
    &+\frac{\kappa}{2}B.&\label{r}
\end{align}

where
\begin{align}
    H'_{MF}&=H_{MF}-\sum_{k,k'\neq0}\eta_{k,k'}(b_k^\dagger b_{k'}+b_{-k}^\dagger b_{-k'})\nonumber\\
        &-\sum_{k,k'\neq0}\gamma_{k,k'}\left[b_{k}b_{-k'}e^{-il(k-k')}+b^\dagger_kb^\dagger_{-k'}e^{il(k-k')}\right]\nonumber\\
        &+\frac{\sqrt{n_0}}{2\sqrt{N}}\frac{\Omega^2}{\Delta}\sum_{k\neq0}e^{-ikl}(u_kb_k-v_kb_{-k}^\dagger)\nonumber\\
        &+\frac{\sqrt{n_0}}{2\sqrt{N}}\frac{\Omega^2}{\Delta}\sum_{k\neq0}e^{ikl}(u_kb_k^\dagger-v_kb_{-k})
\end{align}
is the mean-field Bogoliubov Hamiltonian dressed by the SW Hamiltonian. To treat the coupling between $b_k$ and $r$, as stated, we constrain $r$ to follow the evolution of the bare Bogoliubov quasiparticle fields $b_k$ under an energy pole $s=i\epsilon$. In the regime of narrow measurement bandwidth, this energy $\epsilon$ is set at the dominant and long-lived energy frequency that the overall system is evolving around. Since the tunneling dynamics within the condensate is wide compared to the transition between $a_{j_0}$ and $r$ internal states, and suppose the initial condition of the time evolution is set at the ground-state of the mean-field Hamiltonian $H_{MF}$, we can pick $\epsilon=E_{\text{cond}}$, which is the starting condensate energy, defined in Eq.(\ref{H_MF}). We then Laplace transform Eq.(\ref{r}), and evaluate the (average) value of $r$ in the s-domain at the dominant pole $s = i\epsilon$, to get
\begin{align}\label{AEansatz}
    r&\approx\frac{\kappa/2}{(\Delta'+\epsilon)^2-4\gamma_r^2+\kappa^2/4}\cdot\left\{-i\left[2\gamma_rB^\dagger+(\Delta'+\epsilon)B\right]\right\}\nonumber\\
    &\quad+\frac{\kappa^2/4}{(\Delta'+\epsilon)^2-4\gamma_r^2+\kappa^2/4}B.
\end{align}
Before substituting this approximation for $r$ into Eq.(\ref{bk}), we notice that the term carrying $\gamma_r$ yields 4-th order in $\Omega$ corrections in Eq.(\ref{bk}), which is irrelevant to our perturbation theory. Next, we observe can compact the scalar coefficients next to $r$ and $r^\dagger$ into,
\begin{align}
    -e^{ikl}\frac{\Omega}{\Delta-E_{b,k}}u_k &= [B^\dagger,b_k]\\
    -e^{ikl}\frac{\Omega}{\Delta+E_{b,k}}v_k &=[b_k,B].
\end{align}
To obtain the effective Lindblad jump operator in Eq.(\ref{destroy}), we simply substitute the second term in Eq.(\ref{AEansatz}) into Eq.(\ref{bk}) and force the overall contributions into a Lindblad dissipator form,
\begin{align}\label{ldiss}
    &\mathcal{L}_{\text{diss}}^\dagger(b_k) \approx \mathcal{L}_{\text{diss}}^\dagger[B](b_k)\nonumber\\
    &+\frac{\kappa^2/4}{(\Delta'+\epsilon)^2-4\gamma_r^2+\kappa^2/4}\left([B^\dagger,b_k]B+B^\dagger[b_k,B]\right)\nonumber\\
    &=\left(1+\frac{\kappa^2/4}{(\Delta'+\epsilon)^2-4\gamma_r^2+\kappa^2/4}\right)\mathcal{L}^\dagger_{\text{diss}}[B](b_k)
\end{align}
On the other hand, obtaining the unitary corrections in Eq.(\ref{Heff_loss}) requires a clever re-ordering,
\begin{align}\label{luni}
    &\mathcal{L}_{\text{unitary}}^\dagger(b_k) \approx i[H_\text{new},b_k] - \frac{\kappa}{2}e^{ikj_0}\frac{\Omega}{\Delta - E_{b,k}} u_k r\nonumber\\
    &-\frac{\kappa}{2} r^\dagger e^{ikj_0}\frac{\Omega}{\Delta+E_{b,k}}\nonumber\\
    &=i[H_\text{new},b_k]\nonumber\\
    &-\frac{i\kappa^2/4}{(\Delta'+\epsilon)^2-4\gamma_r^2+\kappa^2/4}\left([B^\dagger,b_k]B - B^\dagger[b_k,B]\right)\nonumber\\
    &=i\left[H_{\text{new}}-\frac{\kappa^2/4}{(\Delta'+\epsilon)^2-4\gamma_r^2+\kappa^2/4}B^\dagger B,b_k\right]
\end{align}
The procedure remains the same for deriving Eq.(\ref{Heff_lossless}) and Eq.(\ref{count}), when the dissipation mechanism is lossless, under the jump operator $\Gamma =\sqrt{\kappa} r^\dagger r$. The only difference is the existence of $\Pi = [B,B^\dagger]$ that arises when the equation of motion for $b_k$ is forced into an effective Lindblad form, as done in Eq.(\ref{ldiss}) and Eq.(\ref{luni}).

\section{Resonant Damping under Modulated Global Driving:}\label{appendix:c}
Firstly, the global drive requires that the bare particles at all sites now have access to the detuned state $\ket{r}$, which means adding to the bare Hamiltonian $H_0$, between the condensate and the uncondensed detuned particles, an additional hopping term,
\begin{align}
    H_0 = H_{MF} - t\sum_j(r_j^\dagger r_{j+1}+\text{h.c}) +\Delta\sum_j r^\dagger_j r_j.
\end{align}
The interaction $V_{\text{all sites}}$ in Eq.(\ref{Vall}) serves as a perturbation, and that allows us to assume that the uncondensed bare particles in the momentum state $r_k$ with dispersion $\mathcal{E}_k=\Delta-2t\cos k$ are weakly coupled to the condensate. Under the SW generator $S_{\text{all sites}}$ in Eq.(\ref{Sall}), in the case of particle loss ($\Gamma = \sqrt{\kappa}r$), the dimensionless factor $\tilde{\kappa}_1(k)$ is given by
\begin{align}
    \tilde{\kappa}_1(k) &= \frac{\kappa/2}{[\tau_{1,k}\tau_{2,k}+\frac{\kappa^2}{4}-\bar{\Omega}^2_k]^2+[\tau_{1,k}-\tau_{2,k}]^2\frac{\kappa^2}{4}}\nonumber\\
    &\cdot\tau_{1,k}\left[\tau_{2,k}^2+\frac{\kappa^2}{4}-\bar{\Omega}^2_k\right],
\end{align}
where,
\begin{align*}
    \bar{\Omega}_k &=2\Omega^2\left(\frac{\mathcal{E}_{k+p}}{\mathcal{E}^2_{k+p}-E_{b,k}^2}+\frac{\mathcal{E}_{k-p}}{\mathcal{E}^2_{k-p}-E_{b,k}^2}\right)u_kv_k\\
    \tau_{1,k}&=\Omega^2\left(\frac{u_k^2}{\mathcal{E}_{k-p}-E_{b,k}}-\frac{v_k^2}{\mathcal{E}_{k+p}+E_{b,k}}\right)\nonumber\\
    &+\mathcal{E}_{k-p}+\epsilon\\
    \tau_{2,k}&=\Omega^2\left(\frac{u_k^2}{\mathcal{E}_{k+p}-E_{b,k}}-\frac{v_k^2}{\mathcal{E}_{k-p}+E_{b,k}}\right)\nonumber\\
    &+\mathcal{E}_{k+p}+\epsilon,
\end{align*}
and $\epsilon$ is the adiabatic elimination energy pole, which is assumed to be the condensate's groundstate energy. In Fig.~\ref{damping_rate}(a), we plot the occupation damping rate $\lambda_k(\Delta)$ near the resonance of each mode, which has the following analytical expression,
\begin{align}
    \lambda_k(\Delta) &= \frac{\kappa}{2}\left[1+\tilde{\kappa}_{1,k}(\Delta)\right]\frac{\Omega^2 u_k^2}{[\mathcal{E}_{k-p}(\Delta)-E_{b,k}]^2}.
\end{align}
On the other hand, when there is no loss ($\Gamma=\sqrt{\kappa}r^\dagger r$), the dimensionless factor $\tilde{\kappa}_2(k)$ is given by
\begin{align}
    &\tilde{\kappa}_2(k) =\nonumber\\
    &\frac{\kappa/2}{(\tau_{1,k}\tau_{2,k}+\frac{\mathcal{K}_{1,k}\mathcal{K}_{1,k}}{4}-\bar{\Omega}_k^2)^2+\frac{1}{4}(\mathcal{K}_{1,k}\tau_{2,k}-\mathcal{K}_{2,k}\tau_{1,k})}\nonumber\\
    &\cdot\left(\tau_{2,k}^2\frac{\mathcal{K}_{1,k}}{2}-\tau_{1,k}\tau_{2,k}\mathcal{K}_{1,k}-\frac{\mathcal{K}_{1,k}\mathcal{K}_{2,k}^2}{8}+\frac{\mathcal{K}_{2,k}}{2}\bar{\Omega}^2_k\right),
\end{align}
where we additionally define the following terms for convenience,
\begin{align*}
    \mathcal{K}_{1,k} &= \left[1+N\Pi_{\text{all sites},k}\right]\kappa\\
    \mathcal{K}_{2,k} &= \left[1+N\Pi_{\text{all sites},-k-p}\right]\kappa.
\end{align*}
Recall that $N$ is the number of site, and $\Pi_{\text{all sites},k}$ arises from the commutator $\Pi_{\text{all sites},k}=[B_k,B^\dagger_k]=\frac{\Omega^2}{(\mathcal{E}_{k-p}-E_{b,k})^2}u_k^2-\frac{\Omega^2}{(\mathcal{E}_{k-p}+E_{b,k})^2}v_k^2$. In Fig.~\ref{damping_rate}(b), we plot the dephasing rate $\eta_k(\Delta)$ near the resonance of each mode, which has the following analytical expression,
\begin{align}
    \eta_k(\Delta) &= \frac{\kappa}{2}\frac{1+\tilde{\kappa}_{1,k}(\Delta)}{\Pi_{\text{all sites},k}(\Delta)}\frac{\Omega^4 u_k^4}{[\mathcal{E}_{k-p}(\Delta)-E_{b,k}]^4}.
\end{align}


\end{document}